\documentclass[10pt,journal,compsoc]{IEEEtran}

\ifCLASSOPTIONcompsoc
  \usepackage[nocompress]{cite}
\else
  \usepackage{cite}
\fi

\ifCLASSINFOpdf
\else
\fi
\usepackage{booktabs} % For formal tables
\usepackage{algorithm, algorithmic}
\usepackage{comment}
\usepackage{enumitem}
\usepackage{amssymb,amsmath}
\usepackage{graphicx}
\usepackage{array}
\usepackage{color}
\usepackage{caption}
\usepackage{graphics}
\excludecomment{comment}

\begin{document}
%
% paper title
% Titles are generally capitalized except for words such as a, an, and, as,
% at, but, by, for, in, nor, of, on, or, the, to and up, which are usually
% not capitalized unless they are the first or last word of the title.
% Linebreaks \\ can be used within to get better formatting as desired.
% Do not put math or special symbols in the title.

\title{Efficient Collaborative Multi-Agent Deep Reinforcement Learning for Large-Scale Fleet Management}
%
%
% author names and IEEE memberships
% note positions of commas and nonbreaking spaces ( ~ ) LaTeX will not break
% a structure at a ~ so this keeps an author's name from being broken across
% two lines.
% use \thanks{} to gain access to the first footnote area
% a separate \thanks must be used for each paragraph as LaTeX2e's \thanks
% was not built to handle multiple paragraphs
%
%
%\IEEEcompsocitemizethanks is a special \thanks that produces the bulleted
% lists the Computer Society journals use for "first footnote" author
% affiliations. Use \IEEEcompsocthanksitem which works much like \item
% for each affiliation group. When not in compsoc mode,
% \IEEEcompsocitemizethanks becomes like \thanks and
% \IEEEcompsocthanksitem becomes a line break with idention. This
% facilitates dual compilation, although admittedly the differences in the
% desired content of \author between the different types of papers makes a
% one-size-fits-all approach a daunting prospect. For instance, compsoc 
% journal papers have the author affiliations above the "Manuscript
% received ..."  text while in non-compsoc journals this is reversed. Sigh.

\author{Kaixiang Lin, Renyu Zhao, Zhe Xu and Jiayu Zhou 

% note the % following the last \IEEEmembership and also \thanks - 
% these prevent an unwanted space from occurring between the last author name
% and the end of the author line. i.e., if you had this:
% 
% \author{....lastname \thanks{...} \thanks{...} }
%                     ^------------^------------^----Do not want these spaces!
%
% a space would be appended to the last name and could cause every name on that
% line to be shifted left slightly. This is one of those "LaTeX things". For
% instance, "\textbf{A} \textbf{B}" will typeset as "A B" not "AB". To get
% "AB" then you have to do: "\textbf{A}\textbf{B}"
% \thanks is no different in this regard, so shield the last } of each \thanks
% that ends a line with a % and do not let a space in before the next \thanks.
% Spaces after \IEEEmembership other than the last one are OK (and needed) as
% you are supposed to have spaces between the names. For what it is worth,
% this is a minor point as most people would not even notice if the said evil
% space somehow managed to creep in.
\IEEEcompsocitemizethanks{\IEEEcompsocthanksitem K. Lin and J. Zhou are with the Department
of Computer Science and Engineering, Michigan State University, East
Lansing, MI, 48823.\protect\\
E-mail: \{linkaixi, jiayuz\}@msu.edu. %\{\} 
\IEEEcompsocthanksitem
R. Zhao and Z. Xu are with Didi Chuxing, Beijing, China.  \protect\\
E-mail: \{zhaorenyu, xuzhejesse\}@didichuxing.com }
\thanks{}}

% The paper headers
% \markboth{Journal of \LaTeX\ Class Files,~Vol.~14, No.~8, August~2015}%
\markboth{Efficient Collaborative MARL for Large-Scale Fleet Management}%
{Shell \MakeLowercase{\textit{et al.}}: Bare Demo of IEEEtran.cls for Computer Society Journals}
% The only time the second header will appear is for the odd numbered pages
% after the title page when using the twoside option.
% 
% *** Note that you probably will NOT want to include the author's ***
% *** name in the headers of peer review papers.                   ***
% You can use \ifCLASSOPTIONpeerreview for conditional compilation here if
% you desire.

% The publisher's ID mark at the bottom of the page is less important with
% Computer Society journal papers as those publications place the marks
% outside of the main text columns and, therefore, unlike regular IEEE
% journals, the available text space is not reduced by their presence.
% If you want to put a publisher's ID mark on the page you can do it like
% this:
%\IEEEpubid{0000--0000/00\$00.00~\copyright~2015 IEEE}
% or like this to get the Computer Society new two part style.
%\IEEEpubid{\makebox[\columnwidth]{\hfill 0000--0000/00/\$00.00~\copyright~2015 IEEE}%
%\hspace{\columnsep}\makebox[\columnwidth]{Published by the IEEE Computer Society\hfill}}
% Remember, if you use this you must call \IEEEpubidadjcol in the second
% column for its text to clear the IEEEpubid mark (Computer Society jorunal
% papers don't need this extra clearance.)

% use for special paper notices
%\IEEEspecialpapernotice{(Invited Paper)}

\newcommand{\lkxcom}[1]{{\color{red}{#1}}}
\newcommand{\lkxn}[1]{{\color{blue}{Notes: #1}}}
\newcommand{\eq}[1]{{Eq~(#1)}}
\newcommand{\va}{{\mathbf{a}}}
\newcommand{\vb}{{\mathbf{b}}}
\newcommand{\vc}{{\mathbf{c}}}
\newcommand{\vd}{{\mathbf{d}}}
\newcommand{\ve}{{\mathbf{e}}}
\newcommand{\vf}{{\mathbf{f}}}
\newcommand{\vg}{{\mathbf{g}}}
\newcommand{\vh}{{\mathbf{h}}}
\newcommand{\vi}{{\mathbf{i}}}
\newcommand{\vj}{{\mathbf{j}}}
\newcommand{\vk}{{\mathbf{k}}}
\newcommand{\vl}{{\mathbf{l}}}
\newcommand{\vm}{{\mathbf{m}}}
\newcommand{\vn}{{\mathbf{n}}}
\newcommand{\vo}{{\mathbf{o}}}
\newcommand{\vp}{{\mathbf{p}}}
\newcommand{\vq}{{\mathbf{q}}}
\newcommand{\vr}{{\mathbf{r}}}
\newcommand{\vs}{{\mathbf{s}}}
\newcommand{\vt}{{\mathbf{t}}}
\newcommand{\vu}{{\mathbf{u}}}
\newcommand{\vv}{{\mathbf{v}}}
\newcommand{\vw}{{\mathbf{w}}}
\newcommand{\vx}{{\mathbf{x}}}
\newcommand{\vy}{{\mathbf{y}}}
\newcommand{\vz}{{\mathbf{z}}}
\newcommand{\valpha}{\bm{\alpha}}

\newcommand{\vA}{{\mathbf{A}}}
\newcommand{\vB}{{\mathbf{B}}}
\newcommand{\vC}{{\mathbf{C}}}
\newcommand{\vD}{{\mathbf{D}}}
\newcommand{\vE}{{\mathbf{E}}}
\newcommand{\vF}{{\mathbf{F}}}
\newcommand{\vG}{{\mathbf{G}}}
\newcommand{\vH}{{\mathbf{H}}}
\newcommand{\vI}{{\mathbf{I}}}
\newcommand{\vJ}{{\mathbf{J}}}
\newcommand{\vK}{{\mathbf{K}}}
\newcommand{\vL}{{\mathbf{L}}}
\newcommand{\vM}{{\mathbf{M}}}
\newcommand{\vN}{{\mathbf{N}}}
\newcommand{\vO}{{\mathbf{O}}}
\newcommand{\vP}{{\mathbf{P}}}
\newcommand{\vQ}{{\mathbf{Q}}}
\newcommand{\vR}{{\mathbf{R}}}
\newcommand{\vS}{{\mathbf{S}}}
\newcommand{\vT}{{\mathbf{T}}}
\newcommand{\vU}{{\mathbf{U}}}
\newcommand{\vV}{{\mathbf{V}}}
\newcommand{\vW}{{\mathbf{W}}}
\newcommand{\vX}{{\mathbf{X}}}
\newcommand{\vY}{{\mathbf{Y}}}
\newcommand{\vZ}{{\mathbf{Z}}}

\newcommand{\cA}{{\mathcal{A}}}
\newcommand{\cB}{{\mathcal{B}}}
\newcommand{\cC}{{\mathcal{C}}}
\newcommand{\cD}{{\mathcal{D}}}
\newcommand{\cE}{{\mathcal{E}}}
\newcommand{\cF}{{\mathcal{F}}}
\newcommand{\cG}{{\mathcal{G}}}
\newcommand{\cH}{{\mathcal{H}}}
\newcommand{\cI}{{\mathcal{I}}}
\newcommand{\cJ}{{\mathcal{J}}}
\newcommand{\cK}{{\mathcal{K}}}
\newcommand{\cL}{{\mathcal{L}}}
\newcommand{\cM}{{\mathcal{M}}}
\newcommand{\cN}{{\mathcal{N}}}
\newcommand{\cO}{{\mathcal{O}}}
\newcommand{\cP}{{\mathcal{P}}}
\newcommand{\cQ}{{\mathcal{Q}}}
\newcommand{\cR}{{\mathcal{R}}}
\newcommand{\cS}{{\mathcal{S}}}
\newcommand{\cT}{{\mathcal{T}}}
\newcommand{\cU}{{\mathcal{U}}}
\newcommand{\cV}{{\mathcal{V}}}
\newcommand{\cW}{{\mathcal{W}}}
\newcommand{\cX}{{\mathcal{X}}}
\newcommand{\cY}{{\mathcal{Y}}}
\newcommand{\cZ}{{\mathcal{Z}}}

\newcommand{\ri}{{\mathrm{i}}}
\newcommand{\rr}{{\mathrm{r}}}

\newcommand{\RR}{\mathbb{R}}
\newcommand{\grad}{{\nabla}}    % gradient
% \usepackage{amsthm}
% \newtheorem*{remark}{Remark}

% for Computer Society papers, we must declare the abstract and index terms
% PRIOR to the title within the \IEEEtitleabstractindextext IEEEtran
% command as these need to go into the title area created by \maketitle.
% As a general rule, do not put math, special symbols or citations
% in the abstract or keywords.
\IEEEtitleabstractindextext{%
\begin{abstract}
Large-scale online ride-sharing platforms have substantially
transformed our lives by reallocating transportation resources to alleviate
traffic congestion and promote transportation efficiency. An efficient fleet
management strategy not only can significantly improve the utilization of
transportation resources but also increase the revenue and customer
satisfaction. It is a challenging task to design an effective fleet
management strategy that can adapt to an environment involving complex
dynamics between demand and supply. Existing studies usually work on a
simplified problem setting that can hardly capture the complicated stochastic
demand-supply variations in high-dimensional space.
In this paper we propose to tackle the large-scale fleet management problem
using reinforcement learning, and propose a contextual multi-agent
reinforcement learning framework including three concrete algorithms to achieve
coordination among a large number of agents adaptive to different
contexts. We show significant improvements of the proposed framework over 
state-of-the-art approaches through extensive empirical studies\footnote{Code can be found at https://github.com/illidanlab/Simulator}.

\end{abstract}

% Note that keywords are not normally used for peerreview papers.
\begin{IEEEkeywords}
Multi-agent Reinforcement Learning; Deep Reinforcement Learning; Fleet Management
\end{IEEEkeywords}}

% make the title area
\maketitle

% To allow for easy dual compilation without having to reenter the
% abstract/keywords data, the \IEEEtitleabstractindextext text will
% not be used in maketitle, but will appear (i.e., to be "transported")
% here as \IEEEdisplaynontitleabstractindextext when the compsoc 
% or transmag modes are not selected <OR> if conference mode is selected 
% - because all conference papers position the abstract like regular
% papers do.
\IEEEdisplaynontitleabstractindextext
% \IEEEdisplaynontitleabstractindextext has no effect when using
% compsoc or transmag under a non-conference mode.

% For peer review papers, you can put extra information on the cover
% page as needed:
% \ifCLASSOPTIONpeerreview
% \begin{center} \bfseries EDICS Category: 3-BBND \end{center}
% \fi
%
% For peerreview papers, this IEEEtran command inserts a page break and
% creates the second title. It will be ignored for other modes.
\IEEEpeerreviewmaketitle

% !TEX ROOT = main.tex
\section{Introduction}
% \IEEEraisesectionheading{\section{Introduction}\label{sec:introduction}}
%% ride share is a significant service. 
Large-scale online ride-sharing platforms such as Uber~\cite{Uber}, Lift~\cite{Lyft}, 
and Didi Chuxing~\cite{Didi} have transformed the way people travel, live and
socialize. By leveraging the advances in and wide adoption of information technologies 
such as cellular networks and global positioning systems,
the ride-sharing platforms redistribute underutilized 
vehicles on the roads to passengers in need of transportation. 
The optimization of transportation resources greatly alleviated traffic congestion
and calibrated the once significant gap between transport demand and supply~\cite{li2016demand}.

%% key challenge is to balance demands and supplies, motivating fleet management 
One key challenge in ride-sharing platforms is to balance the demands 
and supplies, i.e., orders of the passengers and drivers available 
for picking up orders. 
In large cities, although millions of ride-sharing orders are served everyday, 
an enormous number of passengers requests remain unserviced due to the lack 
of available drivers nearby. On the other hand, there are plenty of available
drivers looking for orders in other locations. 
If the available drivers were directed to locations with high demand,  
it will significantly increase the number of orders being served, 
and thus simultaneously benefit all aspects of the society:
utility of transportation capacity will be improved, 
income of drivers and satisfaction of passengers will be increased, 
and market share and revenue of the company will be expanded. 
\emph{fleet management} is a key technical component to
 balance the differences between demand and supply, by 
 reallocating available vehicles ahead of time, to achieve high
efficiency in serving future demand.

%% main motivations for using RL
Even though rich historical demand and supply data are available,
using the data to seek an optimal allocation policy is not an easy 
task. One major issue is that changes in an allocation policy will 
impact future demand-supply, and it is hard for supervised learning 
approaches to capture and model these real-time changes. On the other
hand, the reinforcement learning (RL)~\cite{sutton1998reinforcement}, 
which learns a policy by interacting with a complicated environment, 
has been naturally adopted to tackle the fleet management 
problem~\cite{godfrey2002adaptiveI,godfrey2002adaptiveII,wei2017look}.
However, the high-dimensional and complicated dynamics between demand and supply
can hardly be modeled accurately by traditional RL approaches. 

%% JIAYU NOTE: introduce our problem setting
%%   >> THIS SETTING ITSELF IS NOVEL. <<
Recent years witnessed tremendous success in deep reinforcement learning (DRL) in 
modeling 
intellectual challenging decision-making problems~\cite{mnih2015human,silver2016mastering,silver2017mastering}
that were previously intractable. In the light of such advances, 
in this paper we propose a novel DRL %\emph{multi-agent deep reinforcement learning} 
approach to learn highly efficient allocation policies for fleet management.
%% JIAYU NOTE: challenges in our problem setting. 
% As the first attempt to apply DRL to this problem, there are significant 
% technical challenges we need to complete:
There are significant technical challenges when modeling fleet management using DRL:  

\vspace{+0.02in}
\noindent 1) \emph{Feasibility of problem setting.} 
The RL framework is reward-driven, meaning that a sequence of \emph{actions}
from the policy is evaluated solely by the \emph{reward} signal from environment~\cite{arulkumaran2017brief}. The
definitions of agent, reward and action space are essential for RL. 
If we model the allocation policy using a centralized agent, the action space
can be prohibitively large since an action needs to decide the number of available vehicles
to reposition from each location to its nearby locations. Also, the policy
is subject to a feasibility constraint enforcing that the number of repositioned vehicles 
needs to be no larger than the
current number of available vehicles. To the best of our knowledge, this 
high-dimensional exact-constrain satisfaction policy optimization is not computationally tractable in 
DRL: applying it in a very small-scale problem
could already incur high computational costs~\cite{pham2017optlayer}. 

% \noindent 1) \emph{Complexity of Action Space.} If we model the 
% allocation policy using a centralized agent, the action space
% can be prohibitively large since it needs to decide the number of available vehicles
% to reposition from each location to its nearby locations. Also, the policy
% is subject to a feasibility constraint that the repositioned vehicles needs to be smaller than or equal to
% currently number of available vehicles. To the best of our knowledge, this 
% high-dimensional constrained policy is not computational tractable in multi-agent 
% DRL: applying the constrained policy optimization in a very small scale problem
% already incurs high computational cost~\cite{pham2017optlayer}. 

\vspace{+0.02in}
\noindent 2) \emph{Large-scale Agents.}
% our solution and challenges 
One alternative approach is to instead use a multi-agent DRL setting, where 
each available vehicle is considered as an agent. 
The multi-agent recipe indeed alleviates the curse of dimensionality of action space.
However, such setting creates thousands of agents interacting with the environment at each time. 
Training a large number of agents using DRL is again challenging: 
the environment for each agent is non-stationary since other agents are learning and affecting
the environment at same the time. 
Most of existing studies~\cite{lowe2017multi,foerster2017counterfactual,tampuu2017multiagent} 
allow coordination among only a small set of agents due to high computational costs.

\vspace{+0.02in}
\noindent 3) \emph{Coordinations and Context Dependence of Action space}
Facilitating coordination among large-scale agents remains a challenging task. Since
each agent typically learns its own policy or action-value function that are 
changing over time, it is difficult to coordinate agents for a large number of agents.
Moreover, the action space is dynamic changing 
over time since agents are navigating to different locations and 
the number of feasible actions depends on the geographic context of the location. 

% {\color{red} maybe elaborate a bit more to balance the three challenges.}

% \textbf{Contributions:}
In this paper, we propose a contextual multi-agent DRL framework 
to resolve the aforementioned challenges. Our major contributions are listed as follows:
\begin{itemize}[leftmargin=0.1in]
\item We propose an efficient multi-agent DRL setting for
large-scale fleet management problem by a proper design of agent, reward and
state. 
\item We propose contextual multi-agent reinforcement learning framework in 
which three concrete algorithms: \emph{contextual multi-agent actor-critic} (cA2C), 
\emph{contextual deep Q-learning} (cDQN), and \emph{Contextual multi-agent actor-critic with linear programming} (LP-cA2C) are developed. For the first time in multi-agent DRL, 
the contextual algorithms can not only achieve efficient coordination among thousands of 
learning agents at each time, but also adapt to dynamically changing action spaces. 
% The proposed algorithm is efficient in the sense that coordinations can scale to 
% thousands of agents and it achieve best performance with much less number of repositions.
\item In order to train and evaluate the RL algorithm, we developed a simulator 
that simulates real-world traffic activities perfectly after 
 calibrating the simulator using real historical data provided by Didi Chuxing~\cite{Didi}. 
\item Last but not least, the proposed contextual algorithms significantly outperform 
the state-of-the-art methods in multi-agent DRL with a much less number of repositions needed.
% \vspace{-0.02in}
\end{itemize}

The rest of paper is organized as follows. We first give a literature review
on the related work in Sec~\ref{sec:relatedwork}. Then the problem statement 
is elaborated in Sec~\ref{sec:problem} and the simulation platform we built 
for training and evaluation are introduced in Sec~\ref{sec:simulator}. The methodology 
is described in Sec~\ref{sec:method}. Quantitative and qualitative results are 
presented in Sec~\ref{sec:exp}. Finally, we conclude our work in Sec~\ref{sec:conclusion}.

% summarize previous related work and their limitations to applied in the real scenarios.
%from these perspectives: 
% 1. problem setting different, hard to apply in the real life

% this work, we regard the order dispatching algorithm as one component in
% our environment and focus on reallocating empty vehicles given a fixed 
% order assignment policy,   
% as simultaneously optimizing two policies usually lead to suboptimal performance. 

% 

% % 2. methodology limitaions. 

% It is natural to adopt highly nonlinear function approximation such as neural networks
% to tackle this challenge, which is falls into the filed of deep reinforcement learning and we have not seen any related works. 
% \begin{comment}
% \lkxcom{need solid review; from classical methods to DRL need more motivations}
% \end{comment}
% % previous multi-agent deep reinforcment leanring hard to scale
% \begin{comment}
% \lkxnote{Because the nubmer of idle drivers are larger than orders in real data.
% also the percentage of unserved oders is around $5\%$ averged for the entire city.}
% \end{comment}

% !TEX ROOT = main.tex

\vspace{-0.1in}
\section{Related Works}
\label{sec:relatedwork}
% This work proposed reinforcement learning approaches for dispatch management, and is closely related to the following three areas. 

\noindent\textbf{Intelligent Transportation System.}
Advances in machine learning and traffic data analytics lead to widespread
applications of machine learning techniques to tackle challenging
traffic problems. One trending direction is to incorporate reinforcement
learning algorithms in complicated traffic management problems. There are many
previous studies that have demonstrated the possibility and benefits of reinforcement
learning. Our work has close connections to these studies in terms of problem
setting, methodology and evaluation. 
%
% problem setting
% Bram{~\it{et al}}~\cite{bakker2010traffic} proposed a model-based multi-agent
% Q learning learning framework for traffic light control,  where each
% intersection controller is modeled as an agent and all agents shared a fully
% observable global state. The global action-value function is decomposed into
% the sum of action-value functions of individual junctions, to make the 
% multi-agent system more scalable. 
Among the traffic applications that are closely related to our work, such as taxi dispatch systems or traffic light control algorithms,
multi-agent RL has been explored to model the 
intricate nature of these traffic activities~\cite{bakker2010traffic, seow2010collaborative,maciejewski2013influence}. 
The promising results motivated 
us to use multi-agent modeling in the fleet management problem. %as described in Sec~\ref{sec:problem}.
%
% Both proposed methods in~\cite{seow2010collaborative,maciejewski2013influence}
% tried to leverage communication among agents to improve the dispatch strategy,
% while limited to small number of agents since the explicitly specified actions
% among agents. In our case, more than ten thousands of vehicles are 
%
% Similar to the above setting, we borrow the idea of 
% Since learning a state transition
% function is possible using a vehicle-based
% representation~\cite{wiering2000multi}, a model-based Q-learning framework and three extensions are,
% proposed. To be more specific, the authors decomposed the action-value
% function into the sum of multiple  action-value function of individual
% junctions, which makes the multi-agent system more scalable. 
%
% Methods
In~\cite{godfrey2002adaptiveI}, an adaptive 
dynamic programming approach was proposed to model stochastic dynamic 
resource allocation. It estimates the returns of future states
using a piecewise linear function and delivers actions (assigning orders to vehicles, 
reallocate available vehicles) given states and one step future states values, 
by solving an integer programming problem. In~\cite{godfrey2002adaptiveII}, the authors
further extended the approach to the situations that an action can span across multiple time periods.
These methods are hard to be directly utilized in the real-world setting where orders
can be served through the vehicles located in multiple nearby locations.

% since the dispatch strategy in our case

% we are asked to reallocate the empty vehicles given a dispatch strategy.

% Evaluation
% The evaluation strategies in above works~\cite{seow2010collaborative,maciejewski2013influence} 
% depends on a simulation platform. 

% In this paper~\cite{maciejewski2013influence}:
% \begin{quote}
%  To the best knowledge of the authors, the only application of both microscopic traffic simulation and 
% multi-agent simulation to a real-life scenario has been carried out by Seow et al. in Singapore [11].
% \end{quote}
% Motivation

% The reason to choose simulator as our evaluation.~\cite{abe2004cross,li2010contextual}.
% off-policy evaluation~\cite{precup2000eligibility}

% multi-agent related work. 
\vspace{+0.05in}
\noindent\textbf{Multi-agent reinforcement learning.}
Another relevant research topic is multi-agent reinforcement 
learning~\cite{busoniu2008comprehensive} where a group of agents
share the same environment, in which they receive rewards and
take actions.
% Although the agents
% in the same environment can be fully cooperative, fully competitive
% or in a mixed scenario, we only focus on the cooperative tasks
% since our goal is to manage a set of agents (empty vehicles) to 
% maximize the gross merchandise volume (GMV) of the dispatch platform. 
% the authors discussed using multi-agent's actions as a large joint 
% action and it's able to get optimal solution with Q-learning.
\cite{tan1993multi} compared and contrasted independent 
$Q$-learning and a cooperative counterpart in different settings, and 
empirically showed that the learning speed can benefit from the cooperation among agents.
% cooperative $Q$-learning.
\begin{comment}
% In~\cite{kok2004sparse}, the authors proposed sparse tabular Q-learning, which
% essentially learns two types of Q tables: the independent Q-table that is the same as single 
% agent Q learning without coordination considered, and a sparse Q table that considers the joint actions
% when coordination is needed. 
\end{comment}
Independent $Q$-learning is extended into DRL
in~\cite{tampuu2017multiagent}, where two agents are cooperating or
competing with each other only through the reward. In~\cite{foerster2017counterfactual},
the authors proposed a counterfactual multi-agent policy gradient method that
uses a centralized advantage to estimate whether the action of one agent
would improve the global reward, and decentralized actors to optimize the
agent policy. Ryan~{\it{et al.}} also utilized the framework of decentralized
execution and centralized training to develop multi-agent multi-agent actor-critic
algorithm that can coordinate agents in mixed cooperative-competitive
environments~\cite{lowe2017multi}. However, none of these methods were
applied when there are a large number of agents due to the communication cost among agents. 
Recently, few works~\cite{zheng2017magent,yang2018mean} scaled DRL methods to
a large number of agents, while it is not applicable to apply these methods
to complex real applications such as fleet management. In~\cite{nguyen2017collective,nguyen2017policy}, 
the authors studied large-scale multi-agent planning for fleet management with
explicitly modeling the expected counts of agents. 

% cooperation mechanism among agents haven't been
% explore.

% where the main idea is to implement the parameters-sharing
% DQN and use an agent ID embedding to distinguish agents. However, the authors did not
% include their methods in details and the cooperation mechanism among agents haven't been
% explore. 
% explore the situations where the action space is constantly changing over time, which is essential 
% to the performance in our problem. 

\vspace{+0.05in}
\noindent\textbf{Deep reinforcement learning.}
DRL utilizes neural network function approximations and 
are shown to have largely improved the performance over challenging applications~\cite{silver2017mastering,mnih2015human}.
\begin{comment}
% Deep reinforcement learning approaches used deep neural networks
% to approximate value functions, and were shown to have largely improved
% the performance over traditional reinforcement learning~\cite{silver2016mastering,silver2017mastering,mnih2015human}.
\end{comment}
Many sophisticated DRL algorithms such as DQN~\cite{mnih2015human}, A3C~\cite{mnih2016asynchronous}
% were proposed to adopt the deep learning techniques in various ways and 
were demonstrated to be effective in the tasks in which we have a clear understanding of rules and have easy access to 
millions of samples, such as video games~\cite{brockman2016openai,bellemare2013arcade}. 
However, DRL approaches are rarely seen to be applied in 
complicated real-world applications, especially in those with high-dimensional and non-stationary action space, 
lack of well-defined reward function, and in need of coordination among a large number of agents. 
In this paper, we show that through careful reformulation, the DRL can be applied to tackle the fleet management problem.

% !TEX ROOT = main.tex

\section{Problem Statement}
\label{sec:problem}
\begin{figure}
\centering
\includegraphics[width=0.48\textwidth]{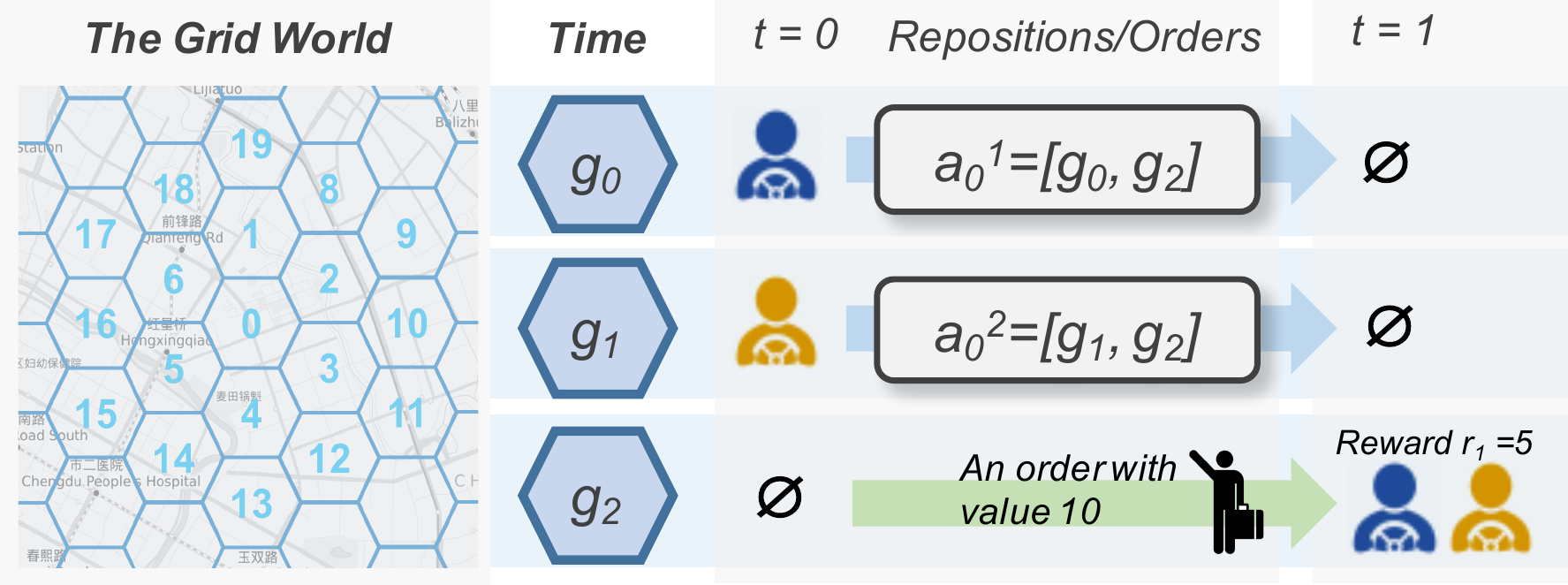}
% \vspace{-0.2in}
\caption{The grid world system and a spatial-temporal illustration of the problem setting. }
\label{fig:problem}
% \vspace{-0.21in}
\end{figure}

% introduce the problem in natural language 
In this paper, we consider the problem of managing a large set of available
homogeneous vehicles for online ride-sharing platforms. 
The goal of the management is to maximize the gross merchandise volume (GMV: the value of all the orders served) 
of the platform by repositioning available vehicles to the locations with larger
demand-supply gap than the current one. This problem belongs to a variant of
the classical fleet management problem~\cite{dejax1987survey}. 
% Once we segment
% the region of interest (e.g., a city) into a set of small regions, the problem is to
% decide how many vehicles we should relocate in each small region at each time span. 
A spatial-temporal illustration of the problem is available in Figure~\ref{fig:problem}.
In this example, we use \emph{hexagonal-grid world} to represent the map
and split the duration of one day into $T=144$ time intervals (one for 10 minutes). 
At each time interval, the orders emerge stochastically in each grid and are served
by the available vehicles in the same grid or six nearby grids. The goal of 
fleet management here is to decide how many available vehicles to relocate 
from each grid to its neighbors in ahead of time, so that most orders can be served. 
% By serving the order,
% the vehicles are moved from one grid and become available in the destination 
% after certain number of time intervals. 
% The vehicles are able to navigate .
% In this work, we do not consider assigning orders to the available vehicles 
% and we encode this procedure into the environment. 
%, elaborated in Sec~\ref{sec:simulator}. 
\begin{comment}
To tackle this problem, we propose to use the reinforcement learning (RL)
framework~\cite{sutton1998reinforcement}, which learns a policy via
iteratively interacting with an environment and optimizing the policy
according to rewards from the environment. There are different ways of modeling
fleet management problem in the RL framework.
% so that serve the stochastic incoming orders more efficiently. 
Ideally, we could use the \emph{single agent learning} setting in RL, 
where we would like to learn a centralized policy that directly 
give optimal allocation action for all grids at each time step. 
The single agent learning uses joint action learners and a single
global reward, and enjoys an optimal solution~\cite{kok2004sparse}. 
However, modeling in this way will lead to a high-dimensional action 
space, and thus incur prohibitive computational costs from extensive exploration.  
\end{comment}

To tackle this problem, we propose to formulate the problem using  
\emph{multi-agent reinforcement learning}~\cite{busoniu2008comprehensive}.
% \lkxcom{is the shared reward function same means reward for all agents same?}
\begin{comment}{This is not
fully cooperative multi-agent learning, since reward is not same for all agents.
Although multiple agents are supposed to coordinate with others to 
maximize the GMV of the platform, which fit into a cooperative task, 
single shared reward for all agents. }
\end{comment}
In this formulation, we use a set of homogeneous agents with small action spaces,
and split the global reward into each grid. This will lead to a much more 
efficient learning procedure than the single agent setting, due to the simplified
action dimension and the explicit credit assignment based on split reward. 
Formally, 
we model the fleet management problem as a Markov game $G$ for $N$ agents, 
which is defined by a tuple $G = (N, \cS, \cA, \cP, \cR, \gamma)$, where 
$N, \cS, \cA, \cP, \cR, \gamma$ are the number of agents, sets of states, joint action space, transition
probability functions, reward functions, and a discount factor respectively. 
The definitions are given as follows:
\begin{itemize}[leftmargin=0.1in]
	\item \textbf{Agent}: 
	We consider an available vehicle (or equivalently an idle driver) as an agent,
	and the vehicles in the same spatial-temporal node are homogeneous, i.e., the
	vehicles located at the same region at the same time interval are
	considered as same agents (where agents have the same policy). %or action-value function.). 
	Although the number of unique heterogeneous agents is always $N$, 
	the number of agents $N_t$ is changing over time.   

	% An action is to move the available vehicle to a
	% nearby grid or stay in the same grid, which gives a set of seven discrete actions. % space of 7 dimensions.
	% For example, the action $a_t = 2$ %$[0, 1, 0, 0, 0, 0, 0]$
	% indicates to relocate the vehicle to the grid indicated by the second
	% index. 
	% We note that alternatively one can consider each grid as an agent
	% and an action is to reallocate available vehicles to nearby grids. Such
	% setting leads to an infinite 7-dimensional actions space, and thus much
	% harder to work with. Furthermore, an action of such is subject to a
	% constraint that it cannot reallocate more vehicles than the available
	% available vehicles in current grid. This needs to be solved using
	% constrained policy gradient
	% methods~\cite{achiam2017constrained,pham2017optlayer}, leading to
	% significant computational challenges and largely decreasing sampling
	% efficiency.
	\begin{comment}
	\lkxnote{For vehicle as agent, at one time step, one grid would generate 7 samples (if the grid
	has six valid neighbors). while for grid as agent, one grid would generate 1 sample.
	}
	\end{comment}
	\item \textbf{State} $\vs_t \in \cS$: %the fully observable global state.
	We maintain a global state $\vs_t$ at each time $t$, considering the
	spatial distributions of available vehicles and orders (i.e. the number of
	available vehicles and orders in each grid) and current time $t$ (using
	one-hot encoding). The state of an agent $i$, $\vs_t^i$, is defined as
	the identification of the grid it located and the shared global state
	i.e. $\vs_t^i = [\vs_t, \vg_j] \in R^{N\times3+T}$, where $\vg_j$ is the one-hot encoding
	of the grid ID. We note that agents located at same grid have the
	same state $\vs_t^i$.

    \item \textbf{Action} $a_t \in \cA = \cA_1 \times...\times \cA_{N_t}$:
	a \emph{joint action} $\va_t=\{a_t^i\}_{1}^{N_t}$ instructing the allocation
	strategy of all available vehicles at time $t$. 
	The action space $\cA_i$ of an individual agent specifies where the 
	agent is able to arrive at the next time, which gives a set of seven 
	discrete actions denoted by $\{k\}_{k=1}^7$. The first six discrete actions
	indicate allocating the agent to one of its six neighboring grids, respectively.
	The last discrete action $a_t^i = 7$ means staying in the current grid.
	For example, the action $a_0^1 = 2$
	means to relocate the $1$st agent from the current grid to the second nearby grid
	at time $0$, as shown in Figure~\ref{fig:problem}. 
	For a concise presentation, we also use $a_t^i \triangleq [\vg_0, \vg_1]$ 
	to represent agent $i$ moving from grid $\vg_0$ to $\vg_1$. 
	% The action of an agent $i$ instructs moving from grid
	% $\vg_0$ to $\vg_1$, and is abbreviated as $a_t^i \triangleq [\vg_0, \vg_1]$. When
	% navigating on a hexagonal map (see Figure~\ref{fig:problem}),
	% the action space is 7: staying in current grid or going to one of its six
	% neighbors. 
	Furthermore, the action space of agents depends on their locations. The agents 
	located at corner grids have a smaller action space.
	We also assume that the action is deterministic: if $a_t^i\triangleq [\vg_0, \vg_1]$, 
	then agent $i$ will arrive at the grid $\vg_1$ at time $t+1$.

	\item \textbf{Reward function} 
	$\cR_{i} \in \cR = \cS \times\cA \rightarrow \RR$: %
	Each agent is associated with a reward function $\cR_{i}$ and all agents in the 
	same location have the same reward function. The $i$-th agent attempts to maximize its own expected 
	discounted return: $\mathbb E\left[\sum\nolimits_{k=0}^{\infty} \gamma^k r_{t+k}^i\right]$.
	The individual reward $r^i_{t}$ for the $i$-th agent associated with the action
	$\va^i_t$ is defined as the averaged revenue of all agents arriving at 
	the same grid as the $i$-th agent at time $t + 1$. Since the individual rewards
	at same time and the same location are same, we denote this reward of agents at time $t$ 
	and grid $\vg_j$ as $r_t(\vg_j)$. %$($R_t(\vg_j)$ as the random variable). 
	Such design of rewards aims at avoiding 
	greedy actions that send too many agents to the location with high value of orders,
	and aligning the maximization of each agent's return with the maximization of 
	GMV (value of all served orders in one day). Its effectiveness is 
	empirically verified in Sec~\ref{sec:exp}.
	\item \textbf{State transition probability} 
	$p(\vs_{t+1}|\vs_{t}, a_t): \cS \times \cA \times \cS \rightarrow [0, 1]$: 
	It gives the probability of transiting to $\vs_{t+1}$ given a joint action $\va_t$ is 
	taken in the current state $\vs_t$. Notice that although the action is deterministic, new
	vehicles and orders will be available at different grids each time, 
	and existing vehicles will become off-line via a random process.
	% learned from real data using maximum likelihood distribution. 

	% \lkxcom{This transition is not deterministic. rewrite problem into a formal SG definition based 
	% on~\cite{littman1994markov}}
\end{itemize}
To be more concrete, we give an example based on the above problem setting
in Figure~\ref{fig:problem}. At time $t=0$, agent $1$
is repositioned from $\vg_0$ to $\vg_2$ by action $a_0^1$, and agent $2$ 
is also repositioned from $\vg_1$ to $\vg_2$ by action $a_0^2$. % respectively.
At time $t=1$, two agents arrive at $\vg_2$, and a new order with value $10$
also emerges at same grid. Therefore, the reward $r_{1}$ for both 
$a_0^1$ and $a_0^2$ is the averaged value received by agents at $\vg_2$, 
which is $10/2 = 5$. 

% This reward setting is only empirically align with the global reward
It's worth to note that this reward design may not lead to the optimal 
reallocation strategy though it empirically leads to good reallocation policy. We give a simple example to 
illustrate this problem. We use the grid world map as show in Figure~\ref{fig:problem}.
At time $t=1$, there is an order with value 100 emerged in $\vg_1$ and another
order with value 10 emerged in $\vg_0$. Suppose we have two agents that are 
available in grid $\vg_0$ at time $t=0$. The optimal reallocation strategy in this case is
to ask one agent stay in $\vg_0$ and another go to $\vg_1$, by which we can 
receive the total reward 110. However, in the current setting, each agent trys to maximize its own 
reward. As a result, both of them will go to $\vg_1$ and receive 50 reward and
none of them will go to $\vg_1$ since the reward they can receive is less than 50. 
However, we show that there are few ways to approximate this global optimal allocation
strategy using the individual action function of each agent.

% !TEX ROOT = main.tex
\section{Contextual Multi-Agent Reinforcement Learning}
\label{sec:method}
In this section, we present two novel contextual multi-agent RL approaches: 
contextual multi-agent actor-critic (cA2C) and 
contextual DQN (cDQN) algorithm. 
% The proposed multi-agent approaches conduct
% explicit coordination on a large number of agents. 
We first briefly introduce the basic 
multi-agent RL method.

\subsection{Independent DQN}
\label{sec:methodiDQN}
% \lkxcom{~\cite{foerster2016learning} this paper introduce independent DQN}
Independent DQN~\cite{tampuu2017multiagent} 
combines independent $Q$-learning~\cite{tan1993multi} and DQN \cite{mnih2015human}.
% It has been applied to the cooperative setting~\cite{tampuu2017multiagent}, where
% two agents have separate deep Q networks. 
% Independent Q-learning~\cite{tan1993multi} extends Q-learning into multi-agent
% setting by having each agent learning its own $Q$ function independently.
% DQN~\cite{mnih2015human} estimates the action-value function through
% minimizing the mean-squared error from Bellman equation and 
A straightforward extension of independent DQN from small scale to
a large number of agents, is to share network parameters and
distinguish agents with their IDs~\cite{zheng2017magent}. The network parameters can be updated 
by minimizing the following loss function, with respect to the transitions collected from all agents:

\begin{equation}
\mathop{\mathbb{E}}%_{\vs_{t}^i, a_t^i, s_{t+1}^i, r_{t+1}^i}
\left[Q(\vs_{t}^i, a_{t}^i; \theta) - \left(r_{t+1}^i + \gamma \max_{a_{t+1}^i} Q(\vs_{t+1}^i, a_{t+1}^i; \theta') \right) \right]^2,
\label{eq:lossdqn}
\end{equation}
where $\theta'$ includes parameters of the target $Q$ network updated periodically, 
and $\theta$ includes parameters of behavior $Q$ network outputting the action value for $\epsilon$-greedy policy,
same as the algorithm described in~\cite{mnih2015human}.
This method could work reasonably well after extensive tunning but it suffers from high variance in 
performance, and it also repositions too many vehicles. 
Moreover, coordination among massive agents is hard to achieve 
since each unique agent executes its action independently based on its action values. 

\subsection{Contextual DQN}
\label{sec:methodcDQN}
Since we assume that the location transition of an agent after the allocation 
action is deterministic, % and the reward corresponds to action $a_t$ is $r_{t+1}$.
% which means the return of agent $i$ doesn't dependents on the $\vs_t$, once $a_t$ is given. 
% Or equivalently it is equal $Q(\vs_t^i, a_t^i)$ can only dependent on $\vs_{t+1}$.
the actions that lead the agents to the same grid should have the same action value.
In this case, the number of unique action-values for all agents should 
be equal to the number of grids $N$. 
Formally, for any agent $i$ where $\vs_t^i = [\vs_t,\vg_i]$, $a_t^i \triangleq [\vg_i, \vg_d]$ and
 $\vg_{i} \in Ner(\vg_{d})$, the following holds:
\begin{align}
Q(\vs_t^i, a_t^i) = Q(\vs_t, \vg_d) %\vE_{\pi}[R_{t+1}(\vg_d) + \gamma G_{t+1}|\vs^i_{t+1}]
\label{eq:cond}
\end{align}
Hence, at each time step, we only need $N$ unique action-values ($ Q(\vs_t, \vg_j), \forall j=1, \dots ,N$) and
the optimization of \eq{\ref{eq:lossdqn}} can be replaced by minimizing the following 
mean-squared loss:
\begin{align}
%L(\theta) = 
\left[
Q(\vs_t, \vg_d; \theta) - \left(r_{t+1}(\vg_d) + \gamma \max_{\vg_{p} \in \text{Ner}(\vg_d)} Q(\vs_{t+1}, \vg_{p};\theta') \right) 
\right]^2.
\label{eq:cdqnl2loss}
\end{align}
This accelerates the learning procedure since the output dimension of the action value function 
is reduced from $\RR^{|\vs_t|}\rightarrow\RR^7 $ to $\RR^{|\vs_t|}\rightarrow\RR$. 
Furthermore, we can build a centralized action-value table at each time for all agents, which
can serve as the foundation for coordinating the actions of agents. 

\noindent\textbf{Geographic context.} 
% When using the hexagonal grid system, the original action space for all agents are 7
% including 6 neighboring grids and the grid it currently located. 
In hexagonal grids systems, border grids and grids surrounded by infeasible grids (e.g.,  
a lake) have reduced action dimensions. 
% For example, the
% agents staying in grid $\vg_{17}$ in Figure~\ref{fig:hexagonal_map} (a) would 
% have feasible action space with four dimensions. 
To accommodate this, 
for each grid we compute a \emph{geographic context} $\vG_{\vg_j} \in \RR^7$,
which is a binary vector that filters out invalid actions for agents
in grid $\vg_j$. 
The $k$th element of vector $\vG_{\vg_j}$ represents the validity of moving 
toward $k$th direction from the grid $\vg_j$. Denote $\vg_d$ as
the grid corresponds to the $k$th direction of grid $\vg_j$, the value of the $k$th element of $\vG_{\vg_j}$ is given by:
\begin{align}
  [\vG_{t, \vg_j}]_k =\left\{
                \begin{array}{ll}
                  1, \hspace{4mm} \text{if } \vg_d  \text{ is valid grid}, \\
                  0, \hspace{4mm} \text{otherwise},
                \end{array}
              \right.
\label{eq:geocontext}
\end{align}
where $k = 0, \dots ,6$ and last dimension of the vector represents direction staying
in same grid, which is always 1. 
% Therefore, the valid logits considering 
% geographic information is $\vq^i *\vG_{\vg_j}$, where $*$ denotes a 
% elementwise multiplication between $\vq^i$ and $\vG_{\vg_j}$. 
\begin{comment}
\lkxnote{The geographic context is fixed once the map is given.}
\end{comment}

\noindent\textbf{Collaborative context.}
% Furthermore, different agents collaborate should be able to collaborate
% in a sense that their actions will not conflict with each other, i.e,
% there are agents moving from grid $\vg_1$ to $\vg_2$ and $\vg_2$ to 
% $\vg_1$ at the same time. 
To avoid the situation that agents are moving in conflict directions (i.e., agents are 
repositioned from grid $\vg_1$ to $\vg_2$ and $\vg_2$ to $\vg_1$ at the same time.),
we provide a \emph{collaborative context} $\vC_{t, \vg_j} \in \RR^7$ for each grid $\vg_j$ at each time.
Based on the centralized action values $Q(\vs_t, \vg_j)$, we restrict
the valid actions such that agents at the grid $\vg_j$ are navigating 
to the neighboring grids with higher action values or staying unmoved.
Therefore, the binary vector $\vC_{t, \vg_j}$ eliminates
actions to grids with lower action values than the action staying unmoved.
Formally, the $k$th element of vector $\vC_{t, \vg_j}$ that corresponds to action value 
$Q(\vs_t, \vg_i)$ is defined as follows:
\begin{align}
  [\vC_{t, \vg_j}]_k =\left\{
                \begin{array}{ll}
                  1, \hspace{4mm} \text{if } Q(\vs_t, \vg_i) >= Q(\vs_t, \vg_j), \\
                  0, \hspace{4mm} \text{otherwise}. 
                \end{array}
              \right.
\label{eq:colcontex}
\end{align}
After computing both collaborative and geographic context, the $\epsilon$-greedy
policy is then performed based on the action values survived from
the two contexts. Suppose the original action values of agent $i$ at time $t$ is $\vQ(\vs_t^i)\in \RR^7_{\geq 0}$, given state $\vs_t^i$, 
the valid action values after applying contexts is as follows:
\begin{align}
\vq(\vs_t^i) = \vQ(\vs_t^i) * \vC_{t, \vg_j} * \vG_{t, \vg_j}.
\label{eq:qvaluemaksing}
\end{align}
The coordination is enabled because that 
the action values of different agents lead to the same location are restricted to be same
so that they can be compared, which is impossible in independent DQN. 
% This method is essentially
% using context to mask the invalid action into zero. 
This method requires that 
action values are always non-negative, which will always hold 
 because that agents always receive nonnegative rewards. 
The algorithm of cDQN is elaborated in Alg~\ref{alg:cDQN}. 

\begin{algorithm}[t]
\caption{$\epsilon$-greedy policy for cDQN}
\label{alg:epsiloncDQN}
\begin{algorithmic}[1]
\REQUIRE Global state $\vs_t$
\STATE Compute centralized action value $Q(\vs_t, \vg_j), \forall j=1,\dots,N$ 
\FOR {$i = 1$ to $N_t$}
\STATE Compute action values $\vQ^i$ by \eq{\ref{eq:cond}}, where $(\vQ^i)_k = Q(\vs_{t}^i, a_{t}^i=k) $.
\STATE Compute contexts $\vC_{t, \vg_j}$ and $\vG_{t, \vg_j}$ for agent $i$.
\STATE Compute valid action values $\vq_{t}^i = \vQ_t^i * \vC_{t, \vg_j} * \vG_{t, \vg_j}$.
\STATE $a_t^i = \text{argmax}_{k} \vq_t^i$ with probability $1 - \epsilon$ otherwise choose 
an action randomly from the valid actions. 
\ENDFOR
\RETURN Joint action $\va_t = \{a_t^i\}_{1}^{N_t}$.
\end{algorithmic}
\end{algorithm}

\begin{algorithm}[t!]
\caption{Contextual Deep Q-learning (cDQN)}
\label{alg:cDQN}
\begin{algorithmic}[1]
\STATE Initialize replay memory $D$ to capacity $M$
\STATE Initialize action-value function with random weights $\theta$ or pre-trained parameters.
\FOR {$m = 1$ to \emph{max-iterations}}
\STATE Reset the environment and reach the initial state $\vs_0$. 
\FOR {$t = 0$ to $T$}
\STATE Sample joint action $\va_t$ using Alg.~\ref{alg:epsiloncDQN}, given $\vs_t$.
\STATE Execute $a_t$ in simulator and observe reward $\vr_t$ and next state $\vs_{t+1}$
\STATE Store the transitions of all agents ($\vs_t^i, a_t^i, r_t^i, \vs_{t+1}^i, \forall i = 1,...,N_t$) in $D$.
\ENDFOR
\FOR {$k = 1$ to $M_1$}
% \STATE Sample a batch of transitions ($\vs_{t_i}^i, a_{t_i}^i, r_{t_i}^i, \vs_{{t_i}+1}^i$) from $D$. 
\STATE Sample a batch of transitions ($\vs_{t}^i, a_{t}^i, r_{t}^i, \vs_{t+1}^i$) from $D$, 
where $t$ can be different in one batch.
% \STATE Compute target $y_{t_i}^i = r_{t_i}^i + \gamma * \max_{a_{t_i+1}^i} Q(\vs_{t_i+1}^i, a_{t_i+1}^i;\theta')$.
\STATE Compute target $y_{t}^i = r_{t}^i + \gamma * \max_{a_{t+1}^i} Q(\vs_{t+1}^i, a_{t+1}^i;\theta')$. 
% $\forall j=1 ,...,b$ for a batch of samples. 
\STATE %Perform a gradient descent on $(y_t^i - Q(\vs_t^i , a_t^i; \theta))^2$ with respect to the 
%network  parameters $\theta$. i.e. 
Update $Q$-network as $\theta \leftarrow \theta + \grad_{\theta} (y_{t}^i - Q(\vs_{t}^i , a_{t}^i; \theta))^2$, 
% where $t$ can be different in the samples of one batch. %denotes the time of transiton agent $i$. %$\forall j=1,\dots,b$.
\ENDFOR
\ENDFOR
\end{algorithmic}
\end{algorithm}

% \subsection{Independent Actor-Critic}
% The independent actor-critic~\cite{foerster2017counterfactual} is a
% simple extension of actor-critic in multi-agent case. 
% \begin{align}
% \grad_{\theta_{p}}J(\theta_p) = \grad_{\theta_{p}} \log\pi_{\theta_p}(a_t^i|\vs_t^i)(r + \gammaV)
% \end{align}
% \subsection{Independent Actor-Critic}
% Independent actor-critic (IAC) is to have multiple agents 
% learn its own policy network and value network independently. 
% In this work, the critic for all agents is based on centralized value
% network and the parameters of policy network are shared among agents.
% The execution remains independent in the sense that multiple agents
% taking their action independently.  An one-step multi-agent 
% advantage actor-critic algorithm can learn it actor and critic by following 
% gradients \eq{\ref{eq:iacpolicy}} and \eq{\ref{eq:iacvalue}}:
% \begin{align}
% \grad_{\theta_p} J(\theta_p)= \grad_{\theta_p}\log(\pi_{\theta_p}(a_t^i|\vs_t^i))(V_{\theta_v}(\vs_t^i) - (r_{t+1}^i + \gamma V_{\theta_v'}(\vs^i_{t+1}))
% \label{eq:iacpolicy}
% \end{align}
% \begin{align}
%  \grad_{\theta_v} L(\theta_v)= \grad_{\theta_v}(V_{\theta_v}(\vs_t^i) - (r_{t+1}^i + \gamma V_{\theta_v'}(\vs^i_{t+1})))^2
% \label{eq:iacvalue}
% \end{align}

\subsection{Contextual Actor-Critic}
\label{sec:methodcA2c}
% We omit the subscript of time $t$ here for the concise of persentation.
We now present the contextual multi-agent actor-critic (cA2C) algorithm,
which is a multi-agent policy gradient algorithm that 
tailors its policy to adapt to the dynamically changing action space. Meanwhile,
it achieves not only a more stable performance but also a much more efficient 
learning procedure in a non-stationary environment.
There are two main ideas in the design of cA2C: 1) A centralized value function shared by all agents
with an expected update; 2) Policy context embedding that establishes explicit coordination
among agents, enables faster training and enjoys the flexibility of regulating policy
to different action spaces. 
% Instead of learning separate critic function for each agent, we directly learn 
The centralized state-value function is learned by minimizing the following loss 
function derived from Bellman equation:
\begin{align}
\label{eq:valueloss}
L(\theta_v) &= (V_{\theta_v}(\vs_t^i) - V_{\text{target}}(\vs_{t+1};\theta_v', \pi))^2,  \\
V_{\text{target}}(\vs_{t+1};\theta_v', \pi) &= \sum\nolimits_{a_t^i}\pi(a_t^i|\vs_t^i) (r_{t+1}^i + \gamma V_{\theta_v'}(\vs^i_{t+1})).
\label{eq:valuetarget}
\end{align}
where we use $\theta_v$ to denote the parameters of the value network and 
$\theta_v'$ to denote the target value network. 
Since agents staying unmoved at the same time are treated homogeneous and share
the same internal state, there are $N$ unique agent states, and thus $N$ unique state-values 
($V(\vs_{t}, \vg_j), \forall j=1,...,N$) at each time. 
The state-value output is denoted by $\vv_t \in \RR^N$, 
where each element $(\vv_t)_j = V(\vs_{t}, \vg_j)$ is the 
expected return received by agent arriving at grid $\vg_j$ on time $t$. In order to
stabilize learning of the value function, we fix a target value network parameterized
by $\theta_v'$, which is updated at the end of each episode.
% The central idea of contextual actor-critic is to embed the context
% derived from this centralized value table to the policy network
% that enable end-to-end training. 
Note that the expected update in~\eq{\ref{eq:valueloss}} and training
actor/critic in an offline fashion are different from the updates in $n$-step actor-critic 
online training using TD error~\cite{mnih2016asynchronous}, 
whereas the expected updates and training paradigm are found to be more stable and sample-efficient. 
This is also in line with prior work in applying actor-critic to real applications~\cite{bahdanau2016actor}.
\begin{comment}
There is something behind. in multi-agent case, does it have some theoretical advantage?
\end{comment}
Furthermore, efficient coordination among multiple agents can be established upon
this centralized value network. 

\noindent\textbf{Policy Context Embedding.}
Coordination is achieved by masking available action space based on the context. 
At each time step, the geographic context is given by~\eq{\ref{eq:geocontext}} and 
the collaborative context is computed according to the value network output:
\begin{align}
  [\vC_{t, \vg_j}]_k =\left\{
                \begin{array}{ll}
                  1, \hspace{4mm} if \ V(\vs_t, \vg_i) >= V(\vs_t, \vg_j), \\
                  0, \hspace{4mm} \text{otherwise}, 
                \end{array}
              \right.
\end{align}
where the $k$th element of vector $\vC_{t, \vg_j}$ 
corresponds to the probability of the $k$th action $\pi(a_t^i=k|\vs_t^i)$. 
Let $\vP(\vs_t^i) \in \RR^7_{> 0}$ denote 
the original logits from the policy network output for the $i$th agent conditioned on state
$\vs_t^i$. Let $\vq_{\text{valid}}(\vs_t^i) = \vP(\vs_t^i) * \vC_{t, \vg_j} * \vG_{\vg_j}$
denote the valid logits considering both geographic and collaborative
context for agent $i$ at grid $\vg_j$,
where $*$ denotes an element-wise multiplication. 
In order to achieve effective masking, we restrict the output logits $\vP(\vs_t^i)$ to be positive.
\begin{comment}
\lkxnote{The geographic context is fixed once the map is given.}
\end{comment}
% The valid logits that considering both geographic and collaborative
% context for agent $i$ at grid $\vg_j$ is $\vq^i_{valid} = \vq^i * \vC_{t, \vg_j} * \vG_{\vg_j}$.
The probability of valid actions for all agents in the grid $\vg_j$ are given by:
\begin{align}
\pi_{\theta_p}(a_t^i=k|\vs_t^i) = [\vq_{valid}(\vs_t^i)]_k = \frac{[\vq_{valid}(\vs_t^i)]_k}{\|\vq_{valid}(\vs_t^i)\|_1 }.
\label{eq:prob}
\end{align}
\begin{comment}
\lkxnote{A practical design on the logits $\vq^i$ is set the activation funciton of last layer 
as relu + 1, which assure all logits are greater than zero and avoid valid logits are all zeros.}
\end{comment}
The gradient of policy can then be written as:
% \begin{align}
% \grad_{\theta_p} J(\theta_p) = \grad_{\theta_p}\log \pi_{\theta_p}(a_t^i|\vs_t^i) A_t(\vs, a_t^i)
% \end{align}
\begin{align}
\grad_{\theta_p} J(\theta_p) = \grad_{\theta_p}\log \pi_{\theta_p}(a_t^i|\vs_t^i) A(\vs_t^i, a_t^i),
\label{eq:policyloss}
\end{align}
where $\theta_p$ denotes the parameters of policy network and 
the advantage $A(\vs_t^i, a_t^i)$ is computed as follows:
\begin{align}
A(\vs_t^i, a_t^i) = r_{t+1}^i + \gamma V_{\theta_v'}(\vs_{t+1}^i) - V_{\theta_v}(\vs_{t}^i).
\label{eq:advantage}
\end{align}
The detailed description of cA2C is summarized in Alg~\ref{alg:cMAAC}. 
\begin{comment}
One main 
difference between is we separate the training of policy network and value network.
\end{comment}

\begin{algorithm}[t!]
\caption{Contextual Multi-agent Actor-Critic Policy forward}
\label{alg:action}
\begin{algorithmic}[1]
\REQUIRE The global state $\vs_t$.
\STATE Compute centralized state-value $\vv_t$%states for all agents $\vs_t^i$, notice that the only state value 
\FOR {i = 1 to $N_t$}
\STATE Compute contexts $\vC_{t, \vg_j}$ and $\vG_{t, \vg_j}$ for agent $i$.
%Compute collaborative context $\vC_{t,\vg_j}$ and geographic context $\vG_{\vg_j}$ for grid $\vg_j$.
\STATE Compute action probability distribution $\vq_{valid}(\vs_t^i)$ for agent $i$ in grid $\vg_j$ as \eq{\ref{eq:prob}}.
\STATE Sample action for agent $i$ in grid $\vg_j$ based on action probability $\vp^i$.
\ENDFOR
\RETURN Joint action $\va_t = \{a_t^i\}_{1}^{N_t}$.
\end{algorithmic}
\end{algorithm}

\begin{figure} 
\centering 
\includegraphics[width=0.48\textwidth]{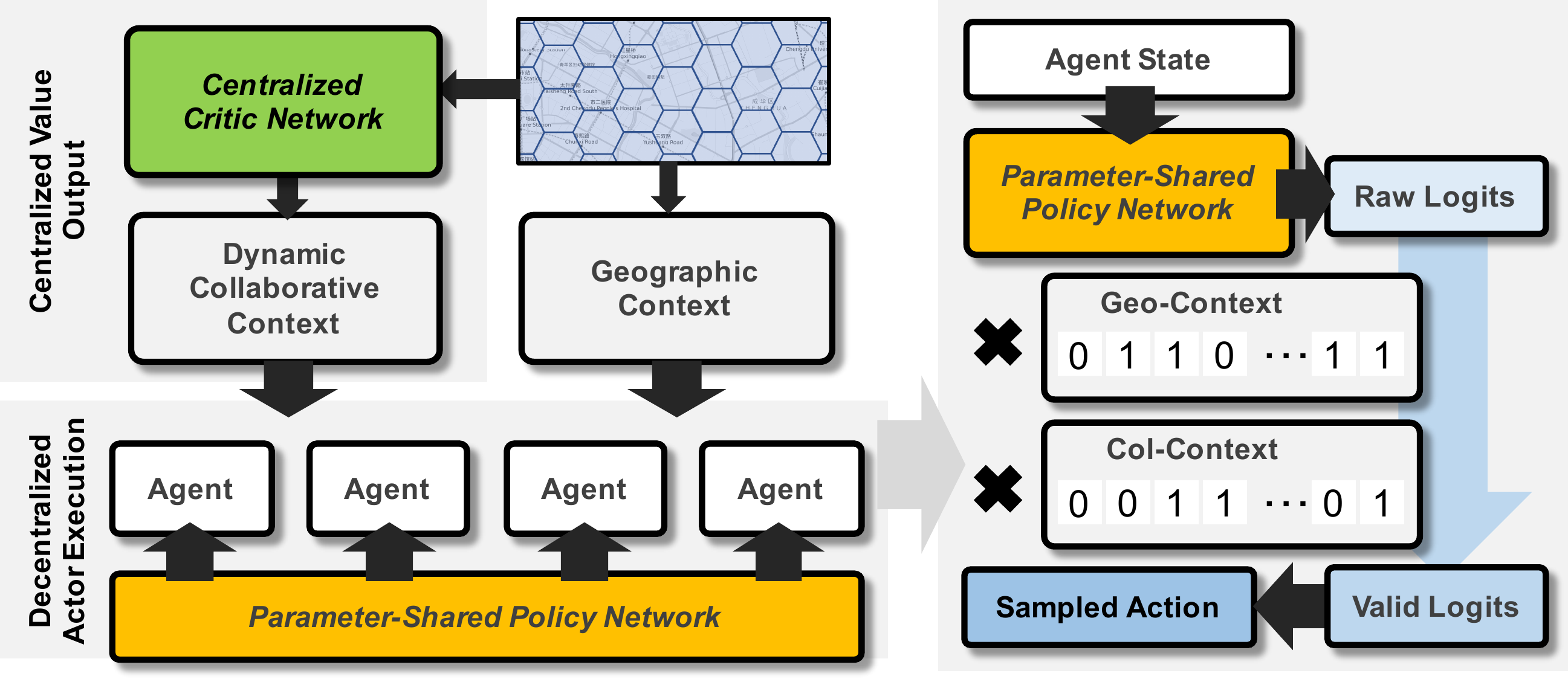}
\caption{Illustration of contextual multi-agent actor-critic. The left
part shows the coordination of decentralized execution based on the output of centralized value network.
The right part illustrates embedding context to policy network. %\lkxcom{use hexagon in the upper plots}
} 
\label{fig:contextualAC} 
% \vspace{-0.1in}
\end{figure}

\begin{algorithm}[h!]
\caption{Contextual Multi-agent Actor-Critic Algorithm for $N$ agents}
\label{alg:cMAAC}
\begin{algorithmic}[1]
% \REQUIRE 
\STATE \textbf{Initialization:}
\STATE Initialize the value network with fixed value table. 
\FOR {$m = 1$ to max-iterations}
\STATE Reset environment, get initial state $\vs_0$. 
% Reset the simulator environment (including bootstrap one days' order, 
% reset initial vehicle distribution, etc.)
\STATE \textbf{Stage 1: Collecting experience}
\FOR {$t = 0$ to $T$}
% \STATE Sample action for all agents based on Alg~\ref{alg:action}.%$a_t^i, i=1,...,N$.
\STATE Sample actions $\va_t$ according to Alg~\ref{alg:action}, given $\vs_t$.
\STATE Execute $\va_t$ in simulator and observe reward $r_t$ and next state $\vs_{t+1}$.
\STATE Compute value network target as~\eq{\ref{eq:valuetarget}} and advantage as~\eq{\ref{eq:advantage}}
 for policy network and store the transitions.
% \STATE Store the transitions in $D$.
\ENDFOR
\STATE \textbf{Stage 2: Updating parameters}
\FOR {$m_1 = 1$ to $M_1$}
\STATE Sample a batch of experience: $\vs_t^i, V_{target}(\vs_{t}^i; \theta_v',\pi)$
\STATE Update value network by minimizing the value loss~\eq{\ref{eq:valueloss}} over the 
batch.
\ENDFOR
\FOR {$m_2 = 1$ to $M_2$}
\STATE Sample a batch of experience:  $\vs_t^i, a_t^i, A(\vs_t^i, a_t^i),\vC_{t,g_j},\vG_{g_j}$,
where $t$ can be different one batch.
\STATE Update policy network as $\theta_p \leftarrow \theta_p + \grad_{\theta_p} J(\theta_p)$.
\ENDFOR
\ENDFOR
\end{algorithmic}
\end{algorithm}

\section{Efficient allocation with linear programming}
\label{sec:lp}
In this section, we present the proposed LP-cA2C that utilizes the 
state value functions learned by cA2C and compute the reallocations
in a centralized view, which achieves the best performance with 
higher efficiency. 

From another perspective, if we formulate this problem as a MDP where we have
a meta-agent  that controls the decisions of all drivers, our goal is to
maximize the long term reward of the platform: $$Q^c(\vs, \va)  =
\vE[\sum_{t=1}^{\infty} \gamma^{t-1}r_t(\vs_t, \va_t) | \vs_0=\vs, \va_0 =
\va, \pi^*].$$ The $\pi^*$ in above formulation denotes the optimal global
reallocation strategy.  Although the sum of immediate reward received by all
agents is equal to the total reward of the platform,  maximizing the long term
reward of each agent is not equal to maximize the long term reward of the
platform, i.e. $\sum_i\max_{a_i} Q(s^i, a^i) \neq \max_{\va}Q^c(\vs, \va)$.
In cooperative multi-agent reinforcement learning, the sum of rewards of
multiple agents is the global reward we want to maximize. In this case, given
a centralized policy ($\pi^*$) for all agents, the summation of long term
reward should be equal to the global long term reward.
\begin{align*}
 \sum_{i=1}^N Q^i(\vs^i, a^i) & = \sum_{i=1}^N E_{\pi^*}\left[\sum_{t=1}^{\infty}\gamma^{t-1} r^i_t \Big| \vs^i_0 =\vs^i,a^i_0= a^i\right] \\
 & = E_{\pi^*}\left[\sum_{t=1}^{\infty}\gamma^{t-1} \sum_{i=1}^N r^i_t \Big| \vs_0 =\vs,\va_0= \va\right]\\
 & = E_{\pi^*}\left[\sum_{t=1}^{\infty}  \gamma^{t-1}r_t\Big|\vs_0 =\vs,\va_0= \va\right]=Q^c(\vs, \va)
\end{align*}
However, in this work, this simple relationship does not hold mainly since the
number of agents ($N_t$) is not static. As shown in~\eq{\ref{eq:reward_diff}},
the global reward at time $t+1$ of the platform is not equal to the sum of all
current agents' reward (i.e. $\sum_{i=1}^{N_t} r^i_{t+1} \neq
\sum_{i=1}^{N_{t+1}} r^i_{t+1} = r_{t+1}$) even given a centralized policy
$\pi^*$.
\begin{align}
\sum_{i=1}^{N_t} Q(\vs^i_t, a^i_t) = \sum_{i=1}^{N_t} \vE_{\pi^*}[ r^i_{t+1} + \gamma \max_{a^i_{t+1}} Q(s^i_{t+1}, a^i_{t+1})] 
  \label{eq:reward_diff}
\end{align}
Ideally, we would like to directly learn the centralized action value function
$Q^c$ while it's computational intractable to explore and optimize the $Q^c$
in the case we have substantially large action space.  Therefore, we need to
leverage the averaged long term reward of each agent to approximate the
maximization of the centralized action-value function $Q^c$. In cDQN, we
approximate this allocation by  avoiding the greedy allocation with
$\epsilon-$greedy strategy even during the evaluation stage. In cA2C,  the
policy will allocate the agents in the same location to its nearby locations
with certain probability according to the  state-values. In fact, we uses this
empirical strategy to better align the joint actions of each individual agent
with the action from optimal reallocation. However, both of the cA2C and cDQN
try to coordinate agents from a localized view, in which each agent only
consider its nearby situation when they are coordinating. Therefore, the
redundant reallocation still exists in those two methods. Other methods that
can approximate the centralized action-value function such as
VDN~\cite{sunehag2017value} and QMIX~\cite{rashid2018qmix} are not able to
scale to large number of agents.

In this work, we propose to approximate the centralized policy by formulating the reallocation 
as a linear programming problem. 
\begin{align}
\label{eq:linearprog}
 \max_{\vy(\vs_t)} \hspace{0.8em}  &  \left( \vv(\vs_t)^T \vA_t - \vc_t^T \right)\vy(\vs_t) - \lambda\| \vD\left(\vo_{t+1} - \vA_t \vy(\vs_t)\right)\|_2^2 \\ %- \sum_i^{N_r} c_i \vY_i + \vV \vI \vY - \lambda(\vo - \vI \vY)^2\\
\text{s.t.} \hspace{0.8em} & \vy(\vs_t) \geq 0 \nonumber \\
           \hspace{0.8em} & \vB_t \vy(\vs_t) = \vd_t \nonumber
\end{align} 
where the vector $\vy(\vs_t) \in R^{N_r(t) \times 1}$ denotes the feasible
repositions for all agents at current  time step $t$. Each element in
$\vy(\vs_t)$ represents one reposition from current grid to its nearby grid.
$N_r(t)$ is the total number of feasible reposition direction. The number of
feasible repositions depends on  the current state values in each grid since
we reallocate agents from location with lower state value to the grid with
higher state value.  $\vA \in R^{N \times N_r(t)}$ is a indicator matrix that
denotes the allocations that dispatch drivers into the grid, i.e. $\vA_{i,j}
\in \{0, 1\}$. $\vA_{i,j}=1$  means the $j$-th reposition reallocates agents
into the $i$-th grid. Similarly,  $\vB \in R^{N \times N_r(t)}$ is the
indicator matrix that denotes the allocations that dispatch drivers out of the
grid. $\vD \in \{0, 1\}^{N \times N}$ is the adjacency matrix denotes the
connectivity of the grid world.  $\vo_{t+1}$ denotes the estimated number of
orders in each grid at next time step. $\vc_t \in R^{N_r(t) \times 1}$ denotes
the cost associated with each reposition and $\vs(\vs_t) \in R^{N \times 1}$
denotes the state value for each grid in time step $t$.

% explain obj 
The first term in \eq{\ref{eq:linearprog}} approximates our goal
that we want to maximize the long term reward of the platform. Since the state
value can be interpreted as the averaged long term reward one agent will
receive  if it appears in certain grid, the first term represents the total
reward minus the total cost associated with the repositions.  However,
optimizing the first term will lead to a greedy solution that reallocates all
the agents to the nearby  grid with highest state value minus the cost. To
alleviate this greedy reallocation, we add the second term to  regularize the
number of agents reallocated to each grid.  Since the agent in current grid
can pick up the orders emerged in nearby grids, we utilize the adjacency
matrix  to regularize the number of agents reallocated into a group of nearby
grids should be close to the number of orders emerged in a group of nearby
grids.  From another point of view, the second term more focus on the
immediate reward since it prefer the solution that allocates right amount of
agents to pick-up the orders without consider the future income that an agent
can receive by that reposition. The regularization parameter $\lambda$ is used
to balance the long term reward and the immediate reward. The two flow
conservation constrains requires the number of repositions should be positive
and the number of repositions from current grid should be equal to the number
of available agents in current grids.

% time complexity. and linear programming. 
Ideally, we need to solve a integer programming problem where our solution
satisfies $\vy(\vs_t) \in \cZ^{N_r}$.  However, solving integer programming is
NP-hard in worst case while solving its linear programming relaxation is in P.
In practice, we solve the linear programming relaxation and round the solution
into integers~\cite{dibangoye2018learning}.

% \input{sec_notes}
% !TEX ROOT = main.tex

\section{Simulator Design}
\label{sec:simulator}
A fundamental challenge of applying RL algorithm in
reality is the learning environment. 
Unlike the standard supervised learning problems where the data is stationary to the
learning algorithms and can be evaluated by the training-testing paradigm,
the interactive nature of RL introduces intricate difficulties on training and evaluation.
%
% One common solution is running the learning algorithm online
% to get the real reward and thus conduct the evaluation. However, 
% it is rarely possible to directly evaluate the algorithm in practice
% due to the expensive cost and large uncertainties in real scenarios,
% especially for the large scale problem such as vehicles reallocation in 
% a city. %<= MOVE THIS TO EVALUTAION IF NEEDED. 
%
One common solution in traffic studies is to build simulators for the environment~\cite{wei2017look,seow2010collaborative,maciejewski2013influence}. 
In this section, we introduce a simulator design that models the 
generation of orders, procedure of assigning orders and key driver 
behaviors such as distributions across the city, on-line/off-line 
status control in the real world. The simulator serves as the training
environment for RL algorithms, as well as their evaluation.  
More importantly, our simulator allows us to calibrate the key 
performance index with the historical data collected from a fleet management system,
and thus the policies learned are well aligned with real-world traffics. 

\noindent\textbf{The Data Description}
The data provided by Didi Chuxing includes orders and trajectories of vehicles 
in two cities including Chengdu and Wuhan. 
% in the center area of a City (Chengdu) in four consecutive weeks. 
Chengdu is covered by a hexagonal grids world consisting of 504 grids.
Wuhan contains more than one thousands grids.
The order information includes order price, origin, destination
and duration. The trajectories contain the positions (latitude and longitude)
and status (on-line, off-line, on-service) of all vehicles every few seconds. 

\begin{figure*}[t!]
% \vspace{-0.1in}
\centering
\includegraphics[width=\textwidth]{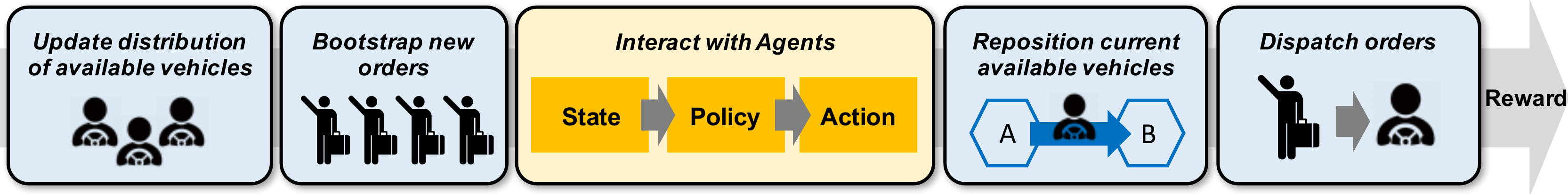}
\caption{Simulator time line in one time step (10 minutes).}
\label{fig:timeline}
% \vspace{-0.2in}
\end{figure*}

\noindent\textbf{Timeline Design.}
% \lkxcom{Add detailed descriptions here. }
% In simulator, each time step represent a 10-minute time interval. 
In one time interval (10 minutes), the main activities are conducted sequentially, also
illustrated in Figure~\ref{fig:timeline}.
\begin{itemize}[leftmargin=0.1in]
\item\emph{Vehicle status updates:} 
Vehicles will be stochastically set offline (i.e., off from service) or online (i.e., start working) following a 
spatiotemporal distribution learned from real data using the maximum likelihood estimation (MLE). 
Other types of vehicle status updates include finishing current service or allocation. 
In other words, if a vehicle is about to finish its service at the current time step,  
or arriving at the dispatched grid, the vehicles are available for taking 
new orders or being repositioned to a new destination.
% The time granularity and spatial granularity are 10 minutes time interval and grid representation,  
% \noindent2) \emph{Order generation: }
\item \emph{Order generation: }
The new orders generated at the current time step are bootstrapped from real orders occurred
in the same time interval. Since the order will naturally reposition vehicles
in a wide range, this procedure keeps the reposition from orders similar to the
real data. 
% \noindent3) \emph{Interact with agents: }
\item \emph{Interact with agents: }
This step computes state as input to fleet management algorithm and applies the allocations for agents.
% \noindent4) \emph{Order assignments:}
\item \emph{Order assignments:}
All available orders are assigned through a two-stage procedure.
In the first stage, the orders in one grid are assigned to the vehicles 
in the same grid. In the second stage, the remaining unfilled orders are 
assigned to the vehicles in its neighboring grids. In reality, the platform dispatches order to
a nearby vehicle within a certain distance, which is approximately the range covered 
by the current grid and its adjacent grids. Therefore, the above two-stage procedure 
is essential to stimulate these real-world activities and the following calibration.
This setting differentiates our problem from the previous fleet management problem setting (i.e., demands
are served by those resources at the same location only.)
and make it impossible to directly apply the classic methods such as adaptive dynamic programming approaches
proposed in~\cite{godfrey2002adaptiveI,godfrey2002adaptiveII}.
\end{itemize}

\begin{figure}[t!]
% \vspace{-0.05in}
\centering 
\captionsetup{justification=centering}
% \begin{tabular}{cc}
\includegraphics[width=0.45\textwidth]{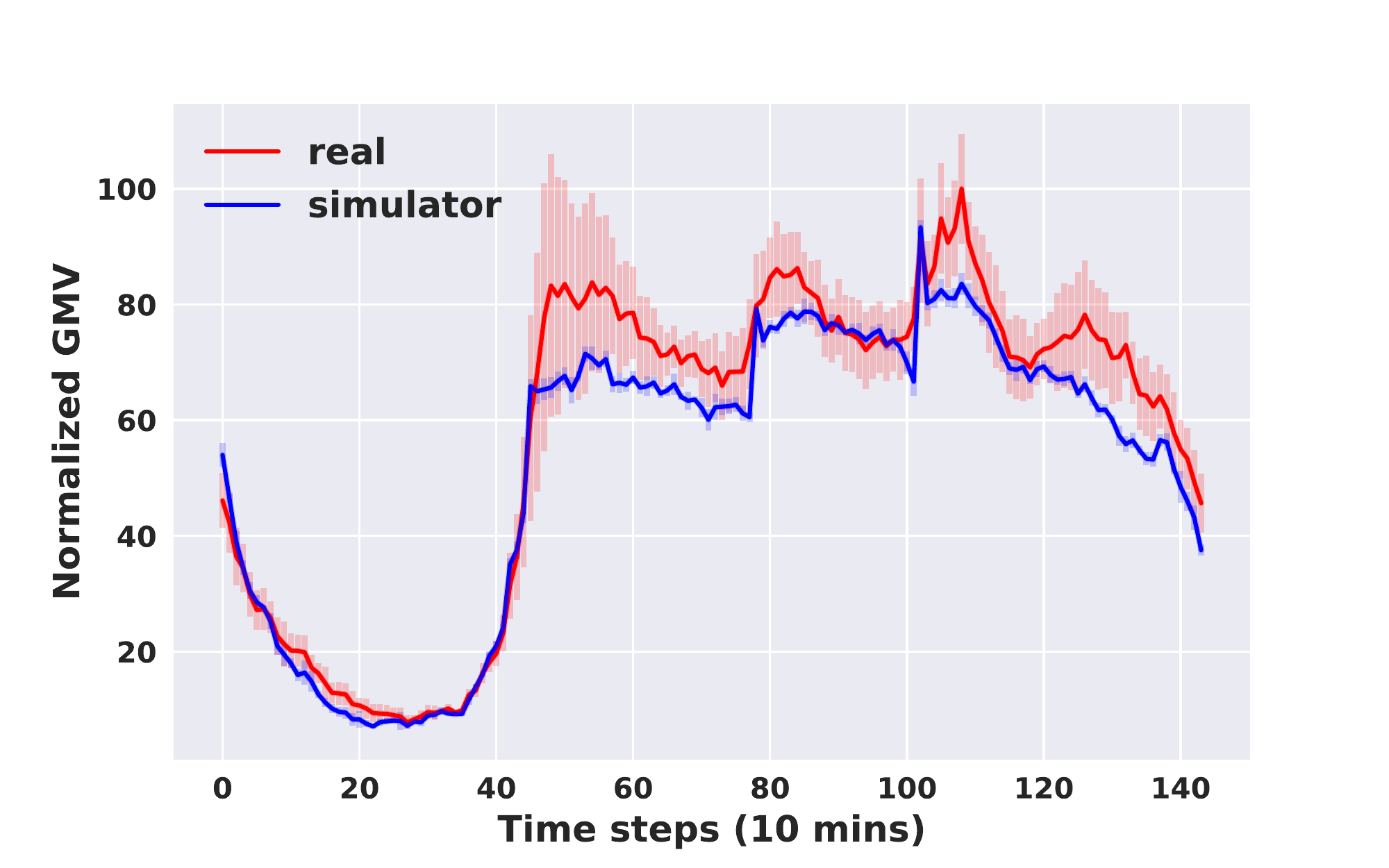}
% &  \includegraphics[width=0.25\textwidth]{fig/order_response_rate7days.pdf} \\
% (a) GMV. % & (b) Order response rate.
% \end{tabular} 
% \vspace{-0.05in}
\caption{The simulator calibration in terms of GMV. 
The red curves plot the GMV values of real data 
averaged over 7 days with standard deviation, in 10-minute time granularity. 
The blue curves are simulated results averaged over 7 episodes.} 
\label{fig:simulatorcalibrate} 
% \vspace{-0.01in}
\end{figure}

\noindent\textbf{Calibration.}
The effectiveness of the simulator is guaranteed by 
calibration against the real data regarding the most important performance measurement: 
the gross merchandise volume (GMV). % and order response rate (ORR). 
As shown in Figure~\ref{fig:simulatorcalibrate}, after the calibration 
procedure, the GMV in the simulator is very similar to that from the ride-sharing platform. 
The $r^2$ between simulated GMV and real GMV is $0.9331$ and
the Pearson correlation is $0.9853$ with $p$-value $p<0.00001$. %$7.226e-104$. 
% Since the real world scenario is complicated and it is not possible to model it perfectly,
% Also, the difference between averaged order response rate in real world and simulator is
% $7.15\%$

% !TEX ROOT = main.tex

\section{Experiments}
\label{sec:exp}

In this section, we conduct extensive experiments to 
evaluate the effectiveness of our proposed method. 

\subsection{Experimental settings}
% Similar as existing works that applied reinforcement learning
% to real applications~\cite{wei2017look,maciejewski2013influence}, 
% we conduct both training and evaluation on the simulator. 
In the following experiments, both of training and evaluation are
conducted on the simulator introduced in Sec~\ref{sec:simulator}. 
For all the competing methods, we prescribe two sets of random seed 
that control the dynamics of the simulator for training and evaluation, respectively.
Examples of dynamics in simulator include order generations, and stochastically status update of all vehicles.
In this setting, we can test the generalization performance of algorithms when it 
encounters unseen dynamics as in real scenarios. 
The performance is measured by GMV (the total value of orders served in the simulator) 
gained by the platform over one episode (144 time steps in the simulator), and 
order response rate (ORR), which is the averaged number of orders served 
divided by the number of orders generated. We use the first 15 episodes 
for training and conduct evaluation on the following ten episodes 
for all learning methods. 
The number of available vehicles at each time in different locations 
is counted by a pre-dispatch procedure. This procedure runs a virtual 
two-stage order dispatching process to compute the remaining available vehicles 
in each location. 
On average, the simulator has 5356 agents per time step waiting for management. 
All the quantitative results of learning methods 
presented in this section are averaged over three runs.

\subsection{Performance comparison}
In this subsection, the performance of following methods
are extensively evaluated by the simulation. 
\begin{itemize}[leftmargin=0.1in]
\item \textbf{Simulation:} This baseline simulates the real scenario without any fleet management.
The simulated results are calibrated with real data in Sec~\ref{sec:simulator}.
\item \textbf{Diffusion:} This method diffuses available vehicles to 
neighboring grids randomly. 
\item \textbf{Rule-based:} This baseline computes a $T \times N$ value table $\vV_{rule}$, where
each element $\vV_{rule}(t, j)$ represents the averaged reward of an agent staying in grid $\vg_j$ 
at time step $t$. The rewards are averaged over ten episodes controlled by random seeds that are different with
testing episodes. With the value table, the agent samples its action based on the
probability mass function normalized from the values of neighboring grids at the next time step.
For example, if an agent located in $\vg_1$ at time $t$ and the current valid actions are 
$[\vg_1, \vg_2]$ and $[\vg_1, \vg_1]$, 
the rule-based method sample its actions from   
$p(a_t^i \triangleq [\vg_1, \vg_j]) = \vV_{rule}(t+1, j)/(\vV_{rule}(t+1, 2)+\vV_{rule}(t+1, 1)), \forall j=1,2$.
\begin{comment} 
random sample action if all value are zeros 
\end{comment}
\item \textbf{Value-Iter:} It dynamically updates the value table based on policy evaluation~\cite{sutton1998reinforcement}. The allocation policy is computed based on the new value table, the same used in the rule-based method, 
while the collaborative context is considered.
\item \textbf{T-$Q$ learning}: The standard independent tabular $Q$-learning~\cite{sutton1998reinforcement} 
learns a table $\vq_{tabular} \in \RR^{T \times N \times 7}$ with $\epsilon$-greedy policy. In this case
the state reduces to time and the location of the agent. 
\item \textbf{T-SARSA}: The independent tabular SARSA~\cite{sutton1998reinforcement} learns
a table $\vq_{sarsa} \in \RR^{T \times N \times 7}$ with same setting of states as T-$Q$ learning. 
\item \textbf{DQN}: The independent DQN is currently the 
state-of-the-art as we introduced in Sec~\ref{sec:methodiDQN}. 
Our $Q$ network is parameterized by a three-layer ELUs~\cite{clevert2015fast} and we adopt the 
$\epsilon$-greedy policy as the agent policy. 
The $\epsilon$ is annealed linearly from 0.5 to 0.1 across the first
15 training episodes and fixed as $\epsilon=0.1$ during the testing. 
% We tune the 
% decay strategy extensively when the under the setting without reposition cost 
% and use it for different settings. 
\begin{comment}
though we observe similar performance with RelU, but it 
\end{comment}
\item \textbf{cDQN}: The contextual DQN as we introduced in Sec~\ref{sec:methodcDQN}. 
The $\epsilon$ is annealed the same as in DQN. At the end of each episode, the $Q$-network
is updated over 4000 batches, i.e. $M_1 = 4000$ in Alg~\ref{alg:cDQN}. To 
ensure a valid context masking, the activation function of the output layer of the $Q$-network
is ReLU + 1.

% \item \textbf{cDQN-$\epsilon$}: 
% \item \textbf{cD-SARSA}:
\item \textbf{cA2C}: The contextual multi-agent actor-critic as we introduced in Sec~\ref{sec:methodcA2c}.
At the end of each episode, both the policy network and the value network 
are updated over 4000 batches, i.e. $M_1 = M_2 = 4000$ in Alg~\ref{alg:cDQN}. Similar to cDQN, 
The output layer of the policy network uses ReLU + 1 as the activation function 
 to ensure that all elements in the original logits $\vP(\vs_t^i)$ are positive. 

\item \textbf{LP-cA2C}: The contextual multi-agent actor-critic with linear programming as 
introduced in Sec~\ref{sec:lp}. During the training state, we use cA2C to explore the
environment and learn the state value function. During the evaluation, we conduct the 
policy given by linear programming. 
\end{itemize}

% \vspace{-0.05in}

\noindent Except for the first baseline, the geographic context is 
considered in all methods 
% by either explicitly casting invalid action-value into zero or encoding into policy (cA2C)
so that the agents will not navigate to the invalid grid.
Unless other specified, the value function approximations and policy network in contextual algorithms
are parameterized by a three-layer ReLU~\cite{he2016identity} with node sizes of 
128, 64 and 32, from the first layer to the third layer. The batch size of all
deep learning methods is fixed as 3000, and we use \textsc{AdamOptimizer} with a learning rate of $1e-3$. 
Since performance of DQN varies a lot when there are a large number of agents, 
the first column in the Table~\ref{tb:results_nocost} for DQN is averaged over the best three runs out of six runs, and the results for all other methods are averaged over three runs.
Also, the centralized critics of cDQN and cA2C are initialized from 
a pre-trained value network using the historical mean of order values computed from 
ten episodes simulation, with different random seeds from both training and evaluation.

To test the robustness of proposed method, we evaluate all competing methods
under different numbers of initial vehicles accross different cities. 
The results are summarized in Table~\ref{tb:results_nocost},~\ref{tb:results_cost},~\ref{tb:results_wuhan}. 
The results of {\it{Diffusion}} improved the
performance a lot in Table~\ref{tb:results_nocost}, possibly because that the method sometimes encourages the available vehicles
to leave the grid with high density of available vehicles, and thus the
imbalanced situation is alleviated. However, in a more realistic setting that we consider 
reposition cost, this method can lead to negative effective due to the highly inefficient 
reallocations. The {\it{Rule-based}} method that 
repositions vehicles to the grids with a higher demand
value, improves the performance of random repositions. The {\it{Value-Iter}} dynamically
updates the value table according to the current policy applied so that it
further promotes the performance upon {\it{Rule-based}}. Comparing the results
of {\it{Value-Iter}}, {\it{T-Q learning}} and {\it{T-SARSA}}, the first method
consistently outperforms the latter two, possibly because that the usage of a centralized value
table enables coordinations, which helps to avoid conflict repositions. 
The above methods simplify the state representation into a spatial-temporal value
representation, whereas the DRL methods account both complex dynamics of supply
and demand using neural network function approximations. As the results shown
in last three rows of Table~\ref{tb:results_nocost},~\ref{tb:results_cost},~\ref{tb:results_wuhan}, the methods with deep learning outperforms the previous one. Furthermore, the contextual algorithms
largely outperform the independent DQN (DQN), which is the state-of-the-art among large-scale multi-agent DRL method and all other competing methods. Last but not least, the lp-cA2C acheive
the best performance in terms of return on investment (the gmv gain per reallocation), GMV, and order
response rate.  

% Tabular Q-learning, tabular SARSA without collaborations.
% Rule-based, Value iter considering collaborations.

\begin{comment}
The performance of DQN is not as stable as our proposed method.
also it is hard to collaborate based on the q value, since the q(s,a)
point to same grid has large variance.
\end{comment}

\begin{table*}
\small
	\centering
	% \captionsetup{justification=centering}
	\caption{Performance comparison of competing methods in terms of GMV and order response rate under the setting that each reposition doesn't incur any cost.
	% The results are averaged over 10 episodes with different random seeds from the training stage. 
	For a fair comparison, the random seeds that control the dynamics of the environment
	are set to be the same across all methods. 
	% The results of all methods are averaged over 3 runs. 
	}
	\label{tab:simulator_performance}
	\vspace{-0.1in}
	\scalebox{0.94}{
	\begin{tabular}{|c|c c|c c|c c|}  \hline
	            & \multicolumn{2}{c|}{$100\%$ initial vehicles}& \multicolumn{2}{c|}{$90\%$ initial vehicles}   &  \multicolumn{2}{c|}{$10\%$ initial vehicles} \\ \hline
	          	& Normalized GMV & ORR        &   Normalized GMV     & ORR    & Normalized GMV    & ORR      \\ \hline\hline
Simulation  	& $100.00\pm0.60$   & $81.80\% \pm 0.37\%$    & $ 98.81 \pm 0.50$    & $80.64\% \pm 0.37\%$   & $ 92.78 \pm 0.79$ & $70.29\% \pm 0.64\%$ \\ \hline
	Diffusion   & $105.68 \pm 0.64$ & $86.48\% \pm 0.54\%$    & $104.44 \pm 0.57$    & $84.93\% \pm 0.49\%$   & $ 99.00 \pm 0.51$ & $74.51\% \pm 0.28\%$ \\ \hline
	Rule-based	& $108.49 \pm 0.40$ & $90.19\% \pm 0.33\%$    & $107.38 \pm 0.55$    & $88.70\% \pm 0.48\%$   & $100.08 \pm 0.50$ & $75.58\% \pm 0.36\%$ \\ \hline
	Value-Iter  & $110.29 \pm 0.70$ & $90.14\% \pm 0.62\%$    & $109.50 \pm 0.68$    & $89.59\% \pm 0.69\%$   & $102.60 \pm 0.61$ & $77.17\% \pm 0.53\%$ \\ \hline
	T-Q learning& $108.78 \pm 0.51$ & $90.06\% \pm 0.38\%$    & $107.71 \pm 0.42$    & $89.11\% \pm 0.42\%$   & $100.07 \pm 0.55$ & $75.57\% \pm 0.40\%$ \\ \hline 
	T-SARSA   	& $109.12 \pm 0.49$ & $90.18\% \pm 0.38\%$    & $107.69 \pm 0.49$    & $88.68\% \pm 0.42\%$   & $ 99.83 \pm 0.50$ & $75.40\% \pm 0.44\%$ \\ \hline
	DQN         & $114.06 \pm 0.66$ & $93.01\% \pm 0.20\%$    & $113.19 \pm 0.60$    & $91.99\% \pm 0.30\%$   & $103.80 \pm 0.96$ & $77.03\% \pm 0.23\%$ \\ \hline\hline
	% IAC         & & & \\ \hline
    cDQN        & $115.19 \pm 0.46$ & $94.77\% \pm 0.32\%$ & \textbf{114.29} $\pm 0.66$ & \textbf{94.00\%} $ \pm 0.53\%$ & $105.29 \pm 0.70$ & $79.28\% \pm 0.58\%$ \\ \hline
	cA2C      & \textbf{115.27} $ \pm 0.70$ & \textbf{94.99\%} $\pm 0.48\%$ & $113.85 \pm 0.69$ & $93.99\% \pm 0.47\%$ &
	          \textbf{105.62} $\pm 0.66$ & \textbf{79.57\%} $\pm 0.51\%$ \\ \hline
	\end{tabular}}
\label{tb:results_nocost}
% \vspace{-0.1in}
\end{table*}
\begin{comment}
\end{comment}

\begin{table*}
% \small
	\centering
	% \vspace{-0.1in}
	% \captionsetup{justification=centering}
	\caption{Performance comparison of competing methods in terms of GMV, order response rate (ORR), and return on invest (ROI) under the setting that each reposition is associated with a cost.}
	\label{tab:cost_performance}
	\vspace{-0.1in}
	% \vspace{-0.1in}
	\scalebox{0.85}{
	\begin{tabular}{|c|c c c|c c c|c c c|}  \hline
	            & \multicolumn{3}{c|}{$100\%$ initial vehicles}& \multicolumn{3}{c|}{$90\%$ initial vehicles}   &  \multicolumn{3}{c|}{$10\%$ initial vehicles} \\ \hline
	          	& Normalized GMV & ORR    & ROI    &   Normalized GMV     & ORR    & ROI& Normalized GMV    & ORR  & ROI    \\ \hline\hline
Simulation  	& $100.00\pm0.60$   & $81.80\% \pm 0.37\%$  & -  & $ 98.81 \pm 0.50$    & $80.64\% \pm 0.37\%$   & - & $ 92.78 \pm 0.79$ & $70.29\% \pm 0.64\%$ & - \\ \hline
	Diffusion   & $103.02 \pm 0.41$ & $86.49\% \pm 0.42\%$  & 0.5890  &$102.35 \pm 0.51$ & $85.00\% \pm 0.47\%$ & 0.7856& $ 97.41 \pm 0.55$ & $74.51\% \pm 0.46\%$ & 1.5600 \\ \hline
	Rule-based	& $106.21 \pm 0.43$ & $90.00\% \pm 0.43\%$ & 1.4868 &$105.30 \pm 0.42$ & $88.58\% \pm 0.37\%$  & 1.7983 &$ 99.37 \pm 0.36$ & $75.83\% \pm 0.48\%$ & 3.2829\\ \hline
	Value-Iter  &   $108.26 \pm 0.65$ & $90.28\% \pm 0.50\%$ & 2.0092 &  $107.69 \pm 0.82$ & $89.53\% \pm 0.56\%$ & 2.5776& $101.56 \pm 0.65$ & $77.11\% \pm 0.44\%$ & 4.5251  \\ \hline
	T-Q learning&   $107.55 \pm 0.58$ & $90.12\% \pm 0.52\%$ & 2.9201 &$106.60 \pm 0.52$ & $89.17\% \pm 0.41\%$  & 4.2052 &$ 99.99 \pm 1.28$ & $75.97\% \pm 0.91\%$  & 5.2527\\ \hline
	T-SARSA   	&   $107.73 \pm 0.46$ & $89.93\% \pm 0.34\%$  & 3.3881 &$106.88 \pm 0.45$ & $88.82\% \pm 0.37\%$  & 5.1559& $ 99.11 \pm 0.40$ & $75.23\% \pm 0.35\%$ &6.8805 \\ \hline
	DQN         & $110.81 \pm 0.68$ & $92.50\% \pm 0.50\%$   &  1.7811 &$110.16 \pm 0.60$ & $91.79\% \pm 0.29\%$  & 2.3790&$103.40 \pm 0.51$ & $77.14\% \pm 0.26\%$ & 4.3770 \\ \hline\hline
	cDQN        & $112.49 \pm 0.42$ & $94.88\% \pm 0.33\%$   & 2.2207 &$112.12 \pm 0.40$ & $94.17\% \pm 0.36\%$  & 2.7708&$104.25 \pm 0.55$ & $79.41\% \pm 0.48\%$ & 4.8340\\ \hline
	cA2C      &   $112.70 \pm 0.64$ & $94.74\% \pm 0.57\%$  & 3.1062 & $112.05 \pm 0.45$ & $93.97\% \pm 0.37\%$  & 3.8085& $104.19 \pm 0.70$ & $79.25\% \pm 0.68\%$ & 5.2124 \\ \hline
	LP-cA2C     & \textbf{113.60} $\pm 0.56$ & \textbf{95.27\%} $\pm 0.36\%$ & \textbf{4.4633}  & \textbf{112.75}$ \pm 0.65$ & \textbf{94.62\%}$ \pm 0.47\%$ & \textbf{5.2719} &\textbf{105.37} $\pm 0.58$ & \textbf{80.15\%} $\pm 0.46\%$& \textbf{7.2949}   \\ \hline
	              %20180919_16-32 20180919_17-02 20180910_19-59 
	\end{tabular}
\label{tb:results_cost}
}
\end{table*}

\begin{table}
% \small
	\centering
	\vspace{-0.1in}
	\caption{Performance comparison of competing methods in terms of GMV, order response rate and return on investment in Wuhan under the setting that each reposition is associated with a cost.
	}
	% \label{tab:cost_wuhan}
	\vspace{-0.1in}
	\scalebox{0.95}{
	\begin{tabular}{|c|c c c|}  \hline
	            
	          	& Normalized GMV & ORR   & ROI     \\ \hline\hline
Simulation  	& $100.00 \pm 0.48$ & $76.56\% \pm 0.45\%$ & - \\ \hline
	Diffusion   & $ 98.84 \pm 0.44$ & $80.07\% \pm 0.24\%$ & -0.2181  \\ \hline
	Rule-based	& $103.84 \pm 0.63$ & $84.91\% \pm 0.25\%$ & 0.5980 \\ \hline
	Value-Iter  & $107.13 \pm 0.70$ & $85.06\% \pm 0.45\%$ & 1.6156\\ \hline
	T-Q learning& $107.10 \pm 0.61$ & $85.28\% \pm 0.28\%$& 1.8302 \\ \hline
	T-SARSA   	& $107.14 \pm 0.64$ & $84.99\% \pm 0.28\%$& 2.0993 \\ \hline
	DQN         & $108.45 \pm 0.62$ & $86.67\% \pm 0.33\%$& 1.0747\\ \hline\hline
	cDQN        & $108.93 \pm 0.57$ & $89.03\% \pm 0.26\%$&  1.1001 \\ \hline
	cA2C      &  $113.31 \pm 0.54$ & $88.57\% \pm 0.45\%$ & 4.4163 \\ \hline
	LP-cA2C    & \textbf{114.92} $\pm 0.65$ & \textbf{89.29\%} $\pm 0.39\%$ &  \textbf{6.1417} \\ \hline
	\end{tabular}
\label{tb:results_wuhan}}
\vspace{-0.1in}
\end{table}

\begin{table}[t!]
\centering
\vspace{-0.02in}
\small
\caption{The effectiveness of contextual multi-agent actor-critic considering dispatch
costs. %The results are averaged over 3 runs. 
}
\vspace{-0.1in}
\scalebox{0.9}{
\begin{tabular}{|c|c|c|c|} \hline
      & Normalized GMV & ORR & Repositions \\ \hline\hline
DQN  & $110.81 \pm 0.68$ & $92.50\% \pm 0.50\%$ & $606932$\\ \hline
cDQN & $112.49 \pm 0.42$ & $94.88\% \pm 0.33\%$ & $562427$ \\ \hline
cA2C &  $112.70 \pm 0.64$ & $94.74\% \pm 0.57\%$ & $408859$ \\ \hline  
LP-cA2C & $113.60 \pm 0.56$ & $95.27\% \pm 0.36\%$ & $304752$ \\ \hline
\end{tabular}}
\label{tb:dispatchcost}
\end{table}

\subsection{On the Efficiency of Reallocations}
In reality, each reposition comes with a cost. In this subsection, we consider
such reposition costs and estimated them by fuel costs. Since the travel
distance from one grid to another is approximately 1.2km and the fuel cost is
around 0.5 RMB/km, we set the cost of each reposition as $c = 0.6$. In this
setting, the definition of agent, state, action and transition probability is
same as we stated in Sec~\ref{sec:problem}. The only difference is that the
repositioning cost is included in the reward when the agent is repositioned
to different locations. Therefore, the GMV of one episode is the sum of all
served order value substracted by the total of reposition cost in one episode.
For example, the objective function for DQN now includes the reposition cost
as follows:
\begin{equation} 
%L(\theta) = 
\mathbb E
\left[Q(\vs_{t}^i, a_{t}^i; \theta) - \left(r_{t+1}^i - c + \gamma \max\nolimits_{a_{t+1}^i} Q(\vs_{t+1}^i, a_{t+1}^i; \theta' \right) 
\right]^2,
\label{eq:costdqn}
\end{equation}
where $a_t^i \triangleq [\vg_o, \vg_d]$, and if $\vg_d=\vg_o$ then $c = 0$, otherwise $c=0.6$. 
Similarly, we can consider the costs in cA2C. However, it is hard 
to apply them to cDQN because that the assumption, that 
different actions that lead to the same location should share the same  
action value, which is not held in this setting. Therefore, instead of 
considering the reposition cost in the objective function, we only
incorporate the reposition cost when we actually conduct our 
policy based on cDQN. Under this setting, the learning objective of 
action value of cDQN is same as in~\eq{\ref{eq:cdqnl2loss}} while
the context embedding is changed from~\eq{\ref{eq:geocontext}} to the
following: 
\begin{align}
  [\vC_{t, \vg_j}]_k =\left\{
                \begin{array}{ll}
                  1, \hspace{4mm} \text{if } Q(\vs_t, \vg_i) >= Q(\vs_t, \vg_j) + c , \\
                  0, \hspace{4mm} \text{otherwise}. 
                \end{array}
              \right.
\label{eq:colcontex_cost}
\end{align}

For LP-cA2C, the cost effect is naturally incorporated in the objective
function as in~\eq{\ref{eq:linearprog}}.  As the results shown in
Table~\ref{tb:dispatchcost},  the DQN tends to reposition more agents while
the contextual algorithms achieve better performance in terms of both GMV and
order response rate, with lower cost. More importantly, the LP-cA2C
outperforms other methods in both of the performance and efficiency. The
reason is that this method formulate the coordination among agents into an
optimization problem, which approximates the maximization of the platform's
long term reward in a centralized version. The centralized optimization
problem can avoid lots of redundant reallocations compared to previous
methods. The training procedures and the network architecture are the same
as described in the previous section.

\begin{figure*}
% \vspace{-0.1in}

\centering
\begin{tabular}{c c}
\hspace{-5mm} \includegraphics[width=0.5\textwidth]{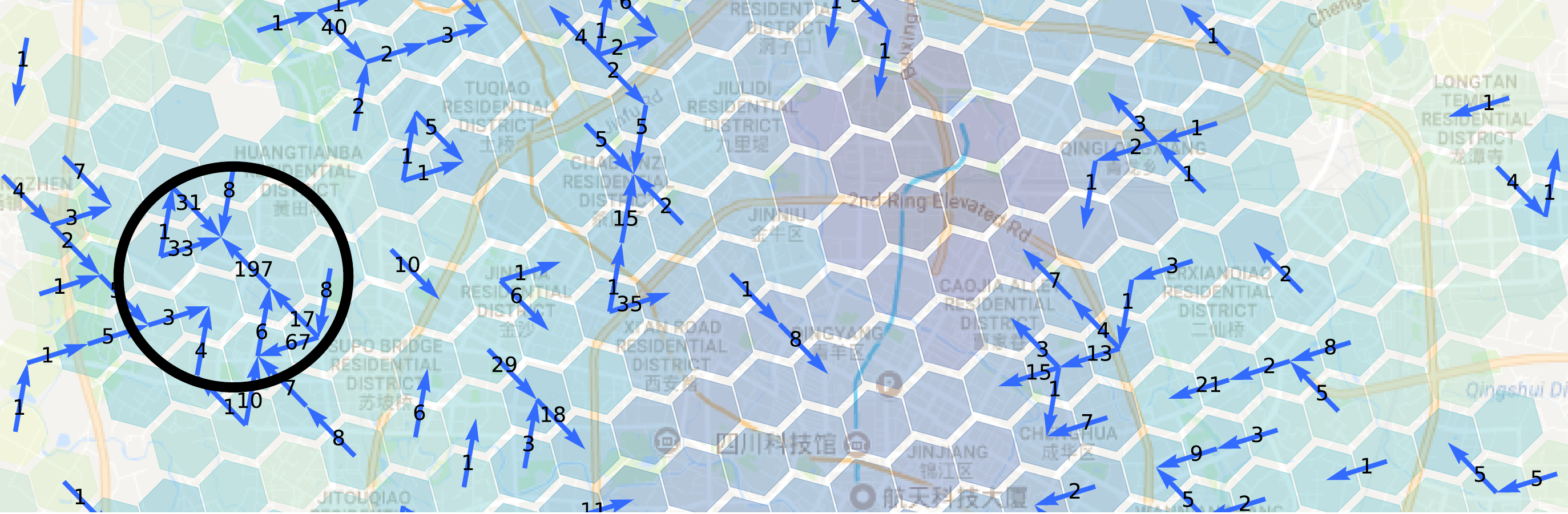}
 & 
\hspace{-4mm} \includegraphics[width=0.5\textwidth]{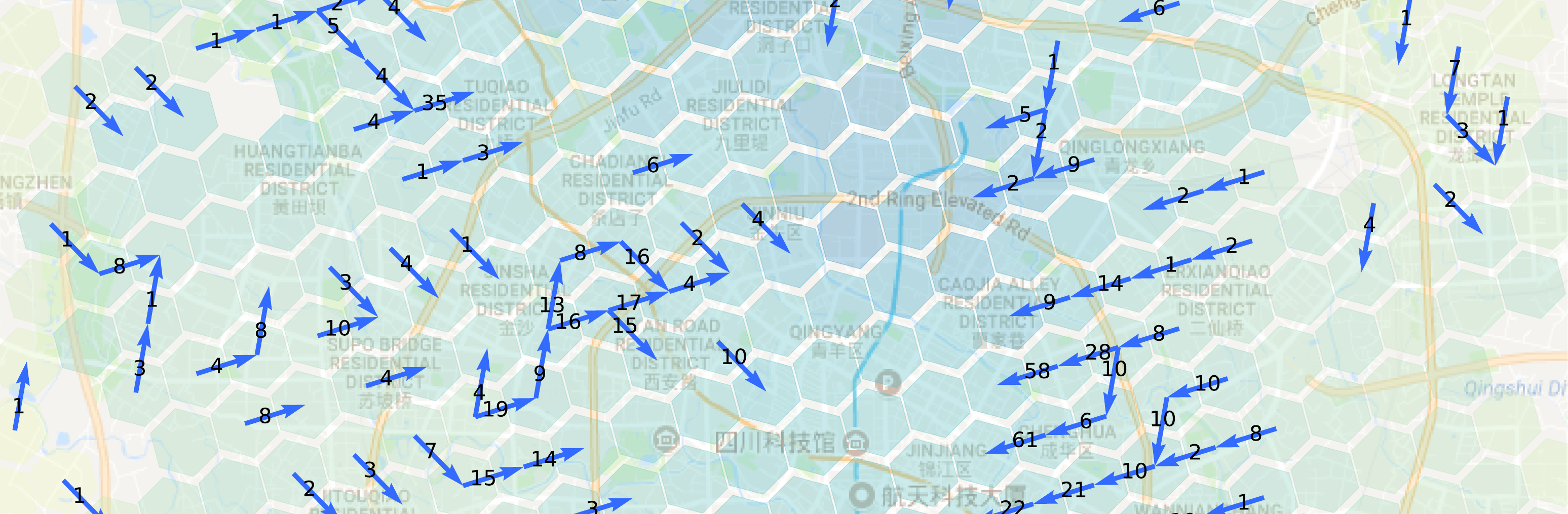}
\\
\hspace{-5mm} \includegraphics[width=0.5\textwidth]{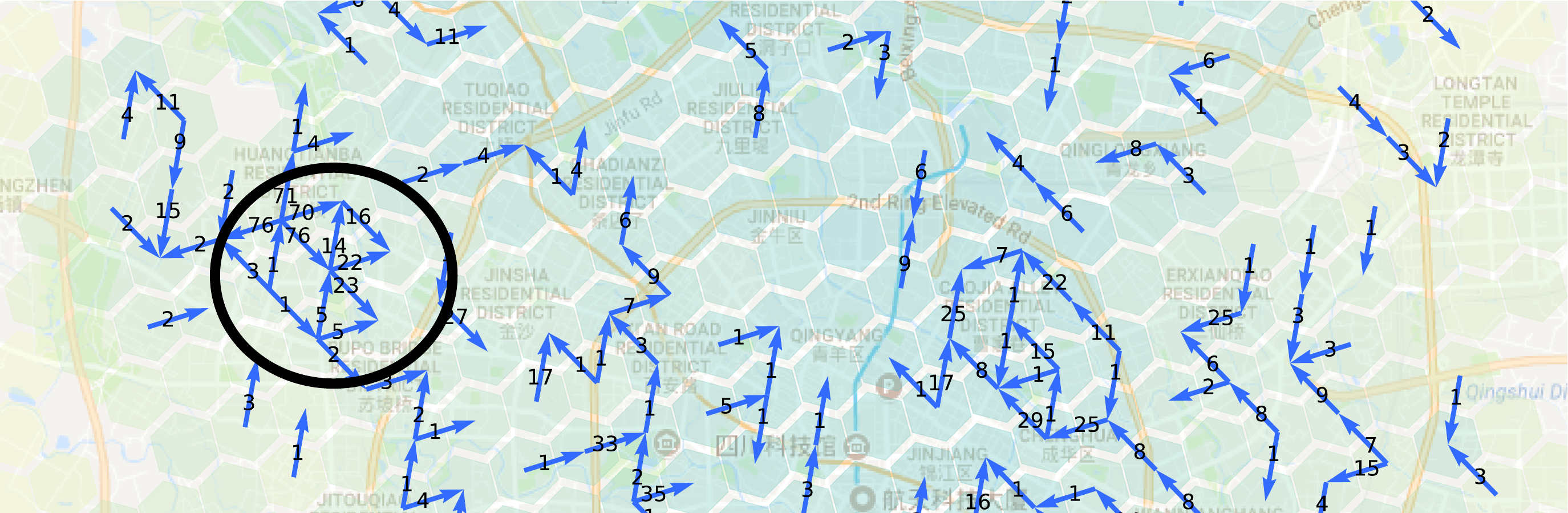}
 & 
\hspace{-4mm} \includegraphics[width=0.5\textwidth]{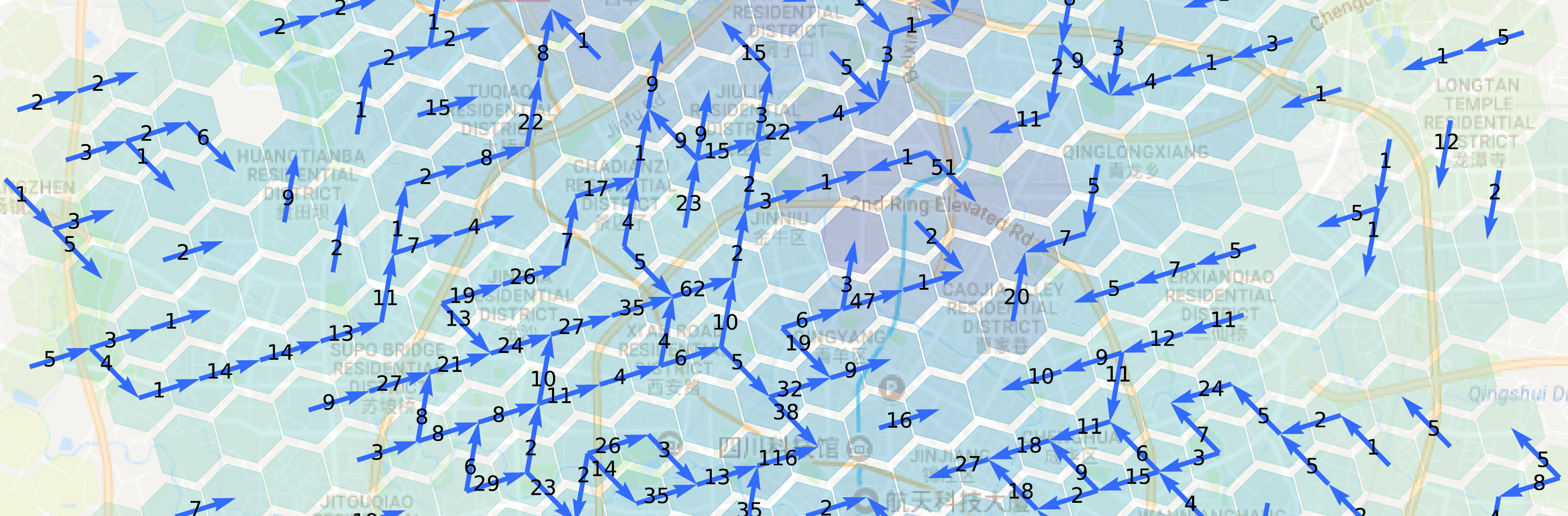}
\\
\hspace{-5mm} \includegraphics[width=0.5\textwidth]{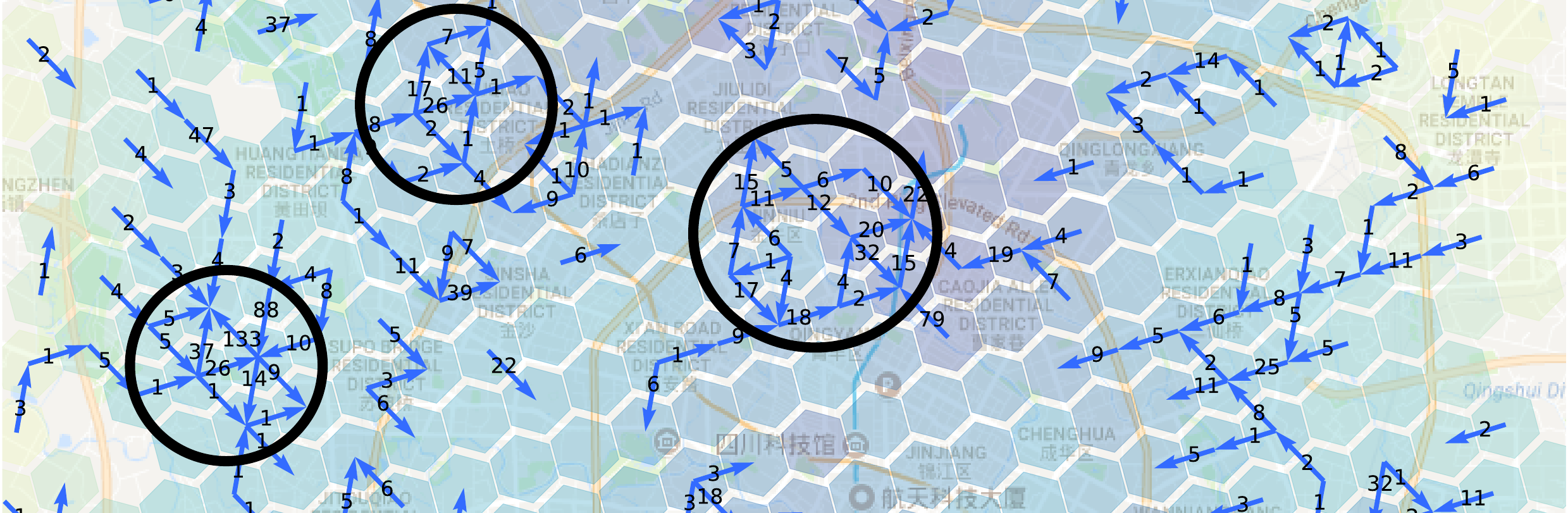}
 & 
\hspace{-4mm} \includegraphics[width=0.5\textwidth]{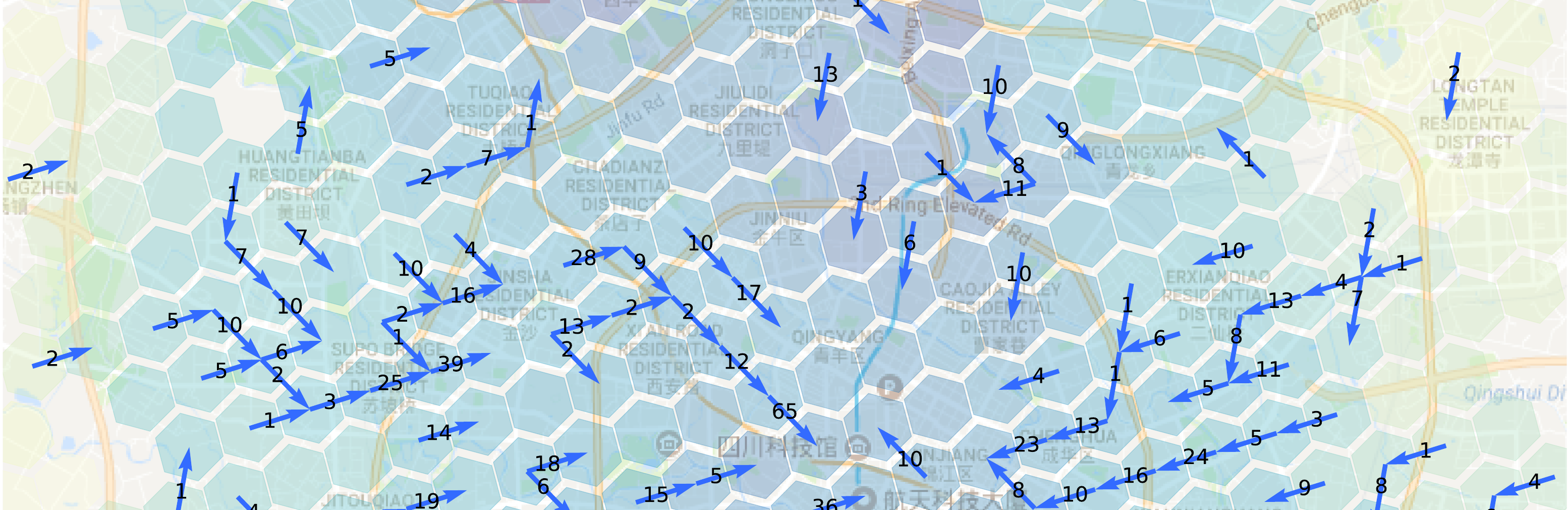}
\\
\hspace{-5mm} \includegraphics[width=0.5\textwidth]{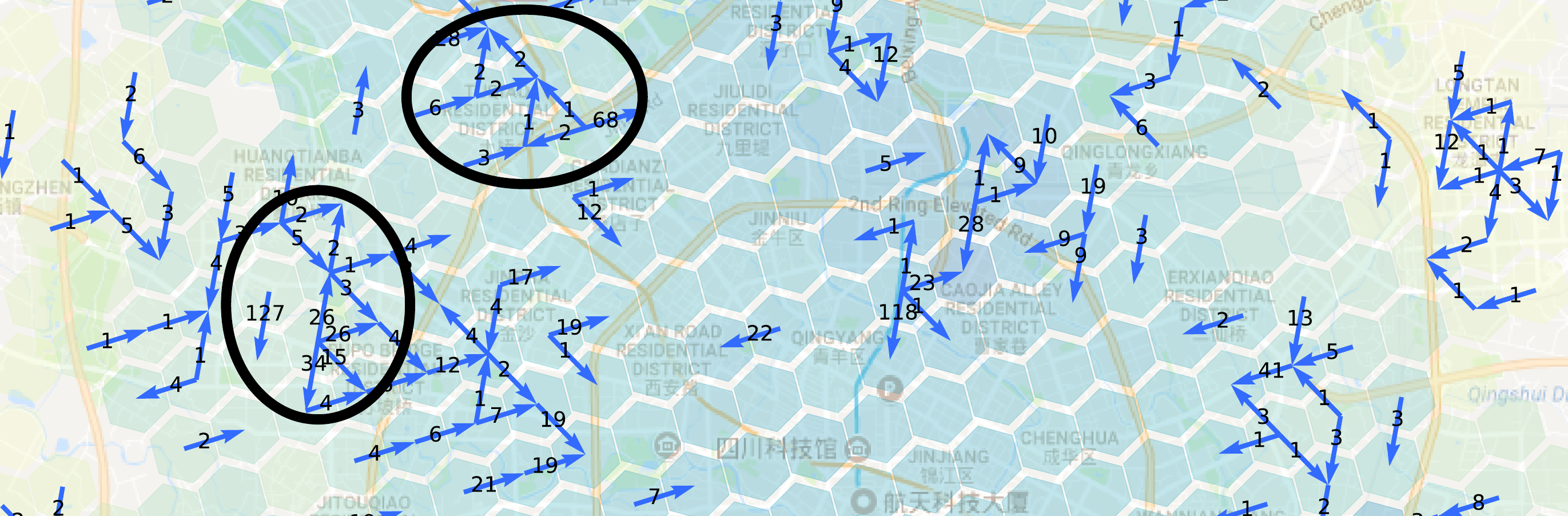}
 & 
\hspace{-4mm} \includegraphics[width=0.5\textwidth]{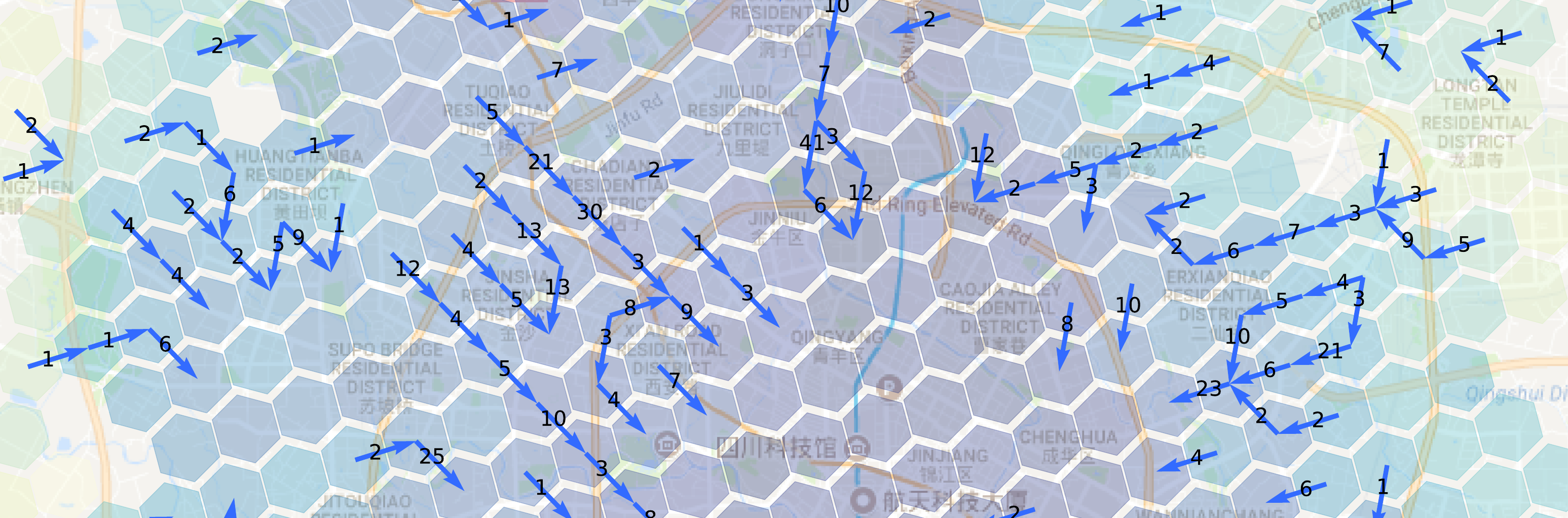}
\\
\hspace{-5mm} \includegraphics[width=0.5\textwidth]{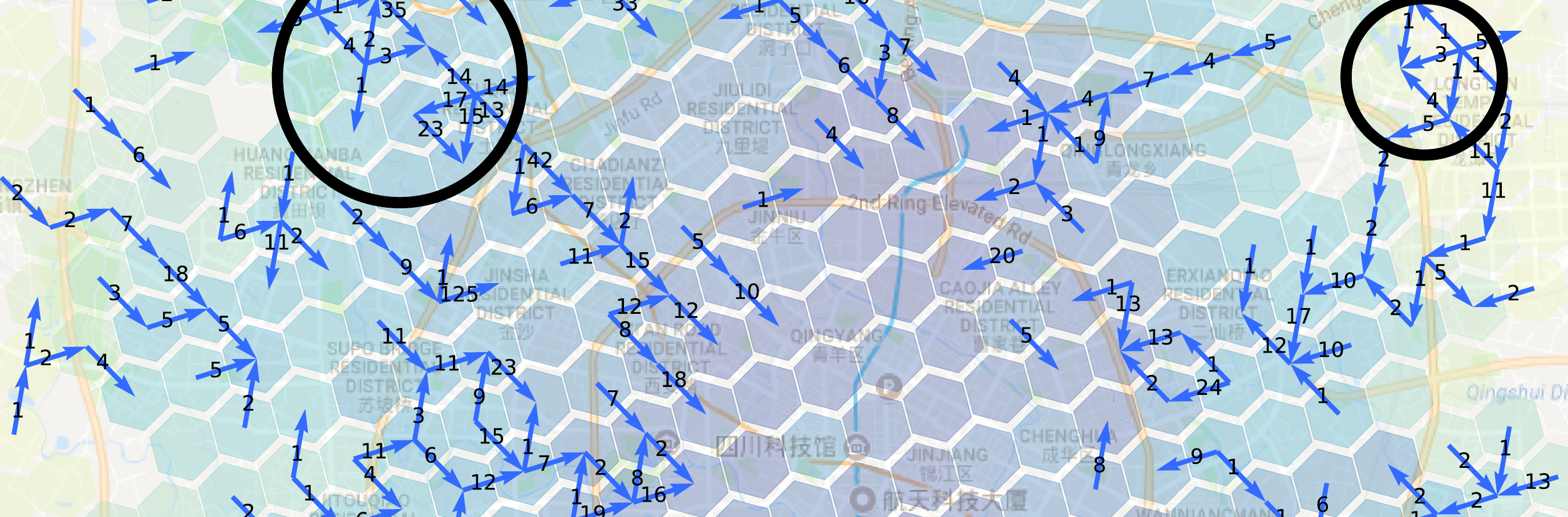}
 & 
\hspace{-4mm} \includegraphics[width=0.5\textwidth]{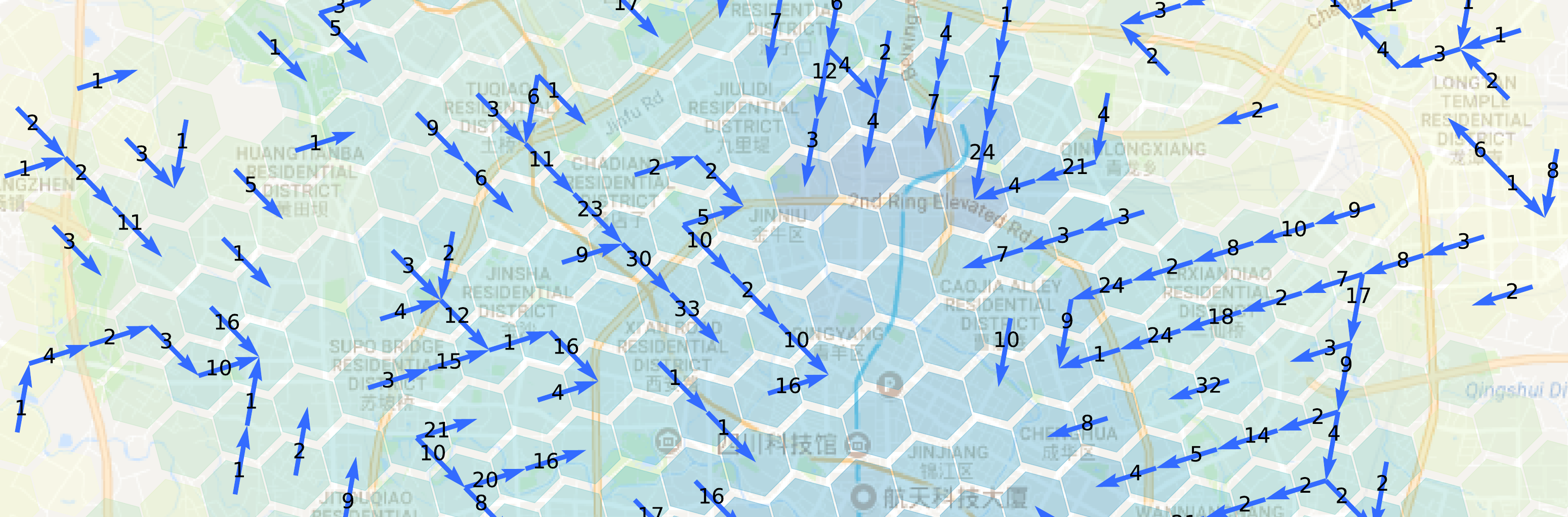}
\\

\hspace{-4mm} (a) cA2C  &  \hspace{-3mm}  (b)  LP-cA2C
\end{tabular}
\caption{The illustration of allocations of cA2C and LP-cA2C at 18:40, 19:40, 20:40, 
21:40, and 22:40, respsectively. The black circles in the left column highlight the 
redundant repositions in the allocation policy given by cA2C. In the right column, there are no such inefficient repositions given by LP-cA2C 
since we compute the repositions in a global view.}
\label{fig:reallcA2CvsLP-cA2C}
% DIDIResearch/data/chengdu/v31_dispatch_ratio1_nogridvalue_withrepnum
% DIDIResearch/data/chengdu/v19_dispatch_ratio1_nogridvalue_withrepnum
% tests/visual_cost_cA2C19.ipynb
% tests/visual_cost_global_cA2C31-casestudy.ipynb
\end{figure*}

To be more concrete, we give a specific scenario to demonstrate that the
efficiency of LP-cA2C. Imaging we would like to ask drivers to move from grid
$A$ to nearby grid $B$ while there is a grid $C$ that is adjacent to both grid
$A$ and $B$. In the previous algorithms, since  the allocation is jointly
given by each agent, it's very likely that we reallocate agents by the short
path $A\rightarrow B$ and longer path $A\rightarrow C \rightarrow B$ when there are sufficient amount of agents can arrive at $B$ from $A$. These inefficient 
reallocations can be avoided by LP-cA2C naturally since the longer path only 
incurs a higher cost which will be the suboptimal solution to our objective 
function compared to the solution only contains the first path. As shown
in Figure~\ref{fig:reallcA2CvsLP-cA2C} (a), 
the allocation computed by cA2C contains
many {\it triangle} repositions as denoted by the black circle, while we didn't
observe these inefficient allocations in Figure~\ref{fig:reallcA2CvsLP-cA2C} (b). 
Therefore, the 
allocation policy delivered by LP-cA2C is more efficient than those given by 
previous algorithms.

% \subsection{Ablation studies}
\subsection{The effectiveness of averaged reward design}
In multi-agent RL, the reward design for each agent 
is essential for the success of learning. In fully cooperative multi-agent
RL, the reward for all agents is a single global reward~\cite{busoniu2008comprehensive},
while it suffers from the credit assignment problem for each agent's action.
Splitting the reward to each agent will alleviate this problem. In this subsection,
we compare two different designs for the reward of each agent: the averaged
reward of a grid as stated in Sec~\ref{sec:problem} and the total reward of a
grid that does not average on the number of available vehicles at that time.
As shown in table~\ref{tb:rewardshaping_col}, the methods with 
averaged reward (cA2C, cDQN) largely outperform those
using total reward, since this design naturally encourages the coordinations
among agents. Using total reward, on the other hand, is likely to reposition an 
excessive number of agents to the location with high demand. 

\begin{table}[t!]
\scriptsize
\centering
\vspace{-0.1in}
\caption{The effectiveness of averaged reward design. The performance of 
methods using the raw reward (second column) is much worse than the performance of
methods using the averaged reward. }
\vspace{-0.1in}
% \begin{tabular}{|c|c|c|}\hline
\begin{tabular}{|>{\centering\arraybackslash}p{.5cm}|>{\centering\arraybackslash}p{3.6cm}|>{\centering\arraybackslash}p{3.7cm}|} \hline
             & Proposed methods & Raw Reward \\ \hline
       	 	 & Normalized GMV/ORR &Normalized GMV/ORR \\ \hline\hline
cA2C  		 & $115.27$$\pm 0.70$/$94.99\% \pm 0.48\%$  & $105.75 \pm 1.17$/$88.09\% \pm 0.74\%$  \\ \hline
cDQN       & $115.19 \pm 0.46$/$94.77\% \pm 0.32\%$ &$108.00\pm0.35$/$89.53\% \pm 0.31\%$ \\ \hline
\end{tabular}
\label{tb:rewardshaping_col}
\end{table}

\begin{figure}
\vspace{-0.1in}
\caption{Convergence comparison of cA2C and its variations without using context
embedding in both settings, with and without reposition costs. The X-axis is
the number of episodes. The left Y-axis denotes the number of conflicts and
the right Y-axis denotes the normalized GMV in one episode.}
\begin{tabular}{c c}
\hspace{-4mm}\includegraphics[width=0.25\textwidth]{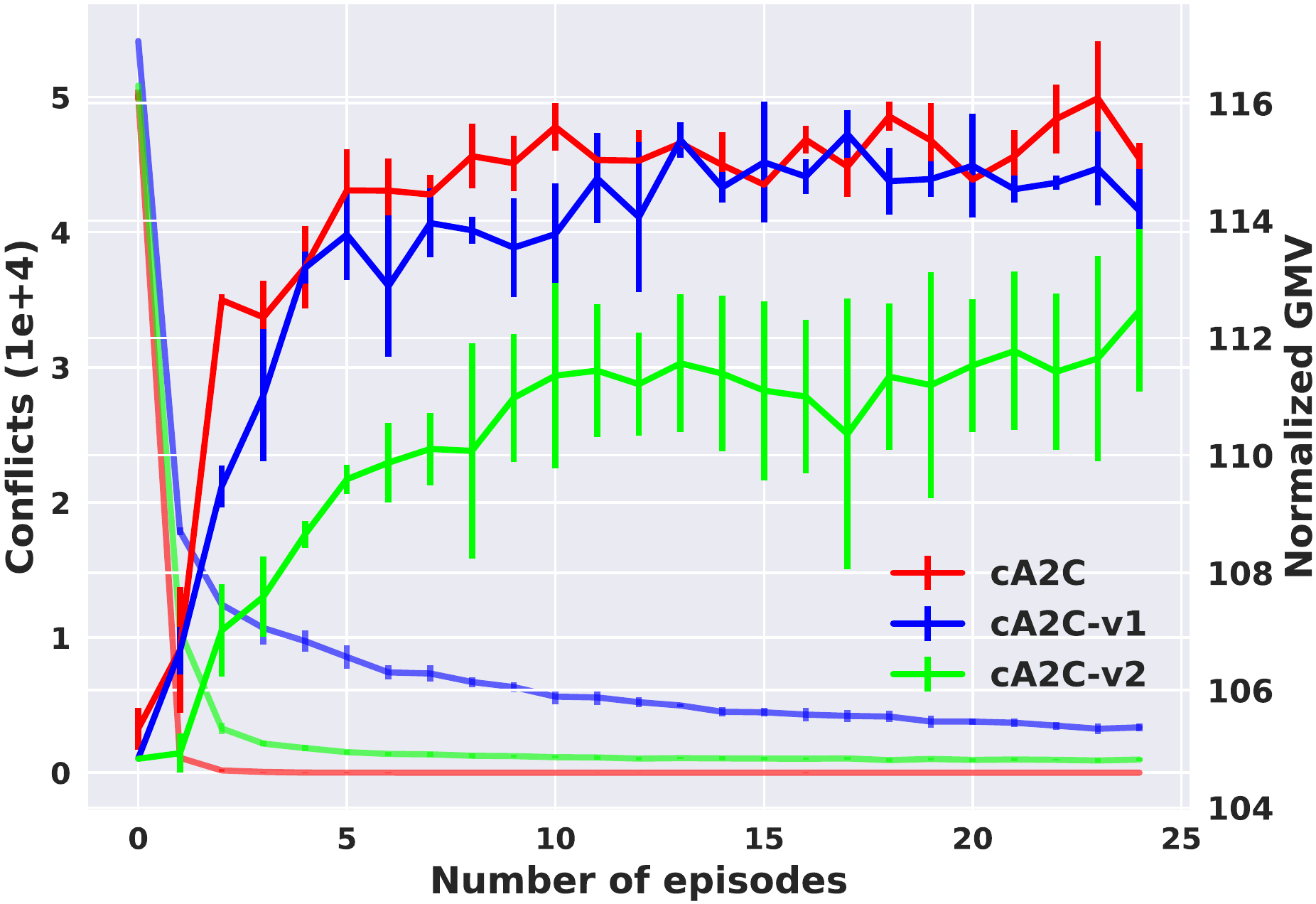} & 
\hspace{-3mm}\includegraphics[width=0.25\textwidth]{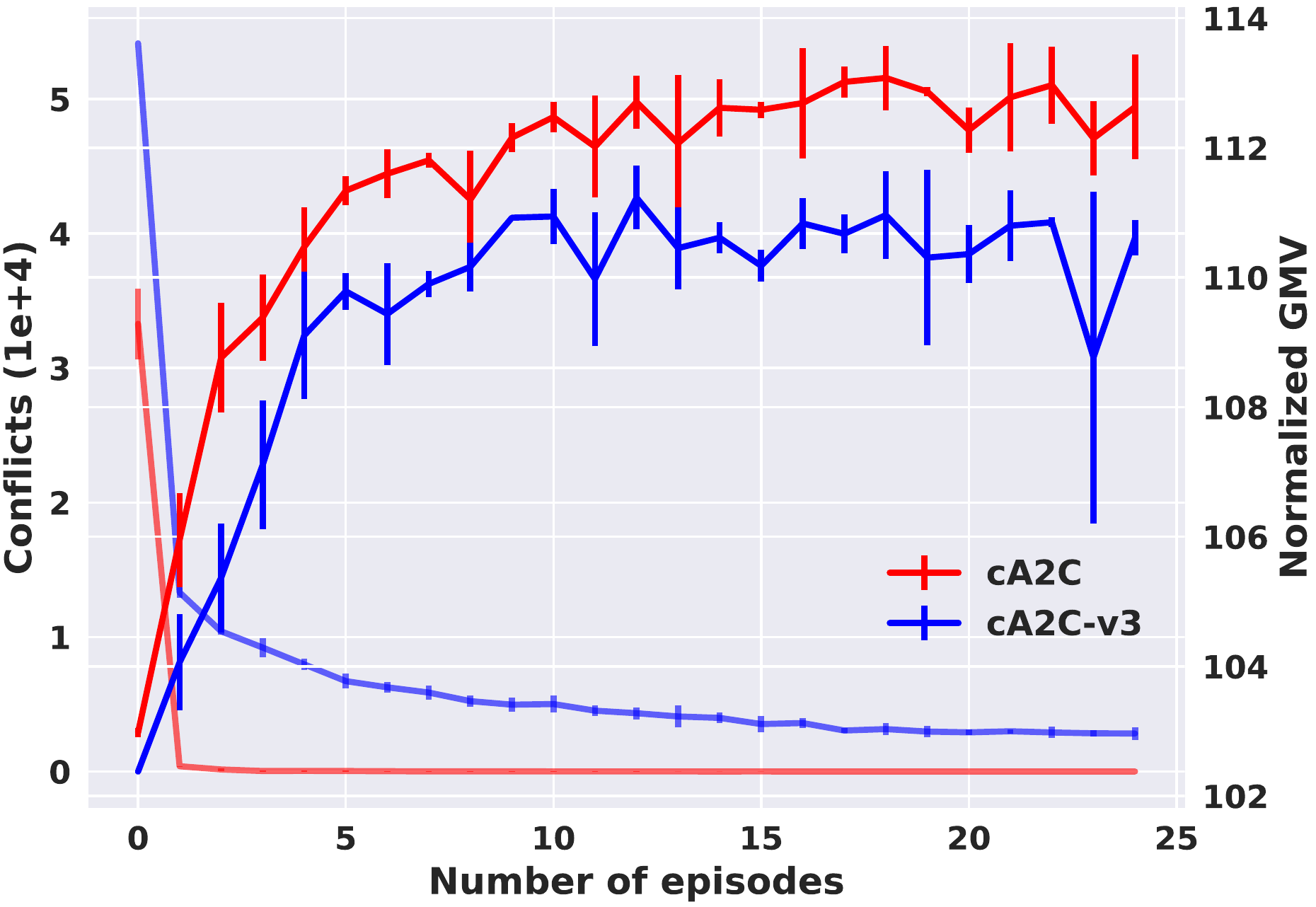}\\
\hspace{-4mm} (a) Without reposition cost  &  \hspace{-3mm}  (b)  With reposition cost
\end{tabular}
\label{fig:contextembed}
% \vspace{-0.2in}
\end{figure}

\begin{table}[t!]
\centering
\caption{The effectiveness of context embedding. 
% The results are averaged over three runs.
}
\label{tb:contextembed}
\vspace{-0.1in}
\small
\begin{tabular}{|c|c|c|} \hline
         & Normalized GMV/ORR & Repositions\\ \hline \hline
         &  \multicolumn{2}{c|}{Without reposition cost} \\ \hline
cA2C     &  $115.27 \pm 0.70$/$94.99\% \pm 0.48\%$ & $460586$\\ \hline
cA2C-v1  &  $114.78 \pm 0.67$/$94.52\% \pm 0.49\%$ & $704568$\\ \hline
cA2C-v2  &  $111.39 \pm 1.65$/$92.12\% \pm 1.03\%$ & $846880$ \\ \hline \hline
		 &  \multicolumn{2}{c|}{With reposition cost} \\ \hline
cA2C     &  $112.70 \pm 0.64$/$94.74\% \pm 0.57\%$ & $408859$ \\ \hline
cA2C-v3  &  $110.43 \pm 1.16$/$93.79\% \pm 0.75\%$ & $593796$\\ \hline %\hline
% cDQN     &  $115.19 \pm 0.46$/$94.77\% \pm 0.32\%$ & $592524$\\ \hline
% cDQN-v1  &  $115.03 \pm 0.60$/$94.83\% \pm 0.43\%$ & $563317$ \\\hline
\end{tabular}
% \vspace{-0.2in}
% from paper_exp_results.py: ablation_context();ablation_compute_cDQN_reallocation()
\end{table}

% \begin{comment}
\subsection{Ablations on policy context embedding}
In this subsection, we evaluate the effectiveness of context embedding,
including explicitly coordinating the actions of different agents through the
collaborative context, and eliminating the invalid actions with geographic
context. 
% We first compare and contrast few variation of cA2C in 
% the settings that considering dispatch cost and without dispatch cost. 
The following variations of proposed methods are investigated in different
settings.
\begin{itemize}
	\item cA2C-v1: This variation drops collaborative context of cA2C
	in the setting that does not consider reposition cost. 
	\item cA2C-v2: This variation drops both geographic and collaborative context
	of cA2C in the setting that does not consider reposition cost. 
	\item cA2C-v3: This variation drops collaborative context of cA2C
	in the setting that considers reposition cost. 
	% \item cDQN-v1: This variation drops collaborative context of cDQN in the 
	% setting that does not considers dispatch cost. 
	% \item In no cost setting: with coordination, converge faster.repositions less drivers. 
	% (average 25 episodes results)
	% \item In no cost setting: repositions less drivers. 
	% \item in the dispatch cost setting: performance much better.  
	% \item cDQN: In no cost setting: with coordination, converge faster. repositions less drivers. 
\end{itemize}

The results of above variations are summarized in Table~\ref{tb:contextembed}
and Figure~\ref{fig:contextembed}.   As seen in the first two rows of
Table~\ref{tb:contextembed} and the red/blue curves in
Figure~\ref{fig:contextembed} (a), in the setting of zero reposition cost, cA2C
achieves the best performance with much less repositions ($65.37\%$) comparing with
cA2C-v1. Furthermore, collaborative context embedding achieves significant
advantages when the reposition cost is considered, as shown in the last
two rows in Table~\ref{tb:contextembed} and Figure~\ref{fig:contextembed} (b). It
not only greatly improves the performance but also accelerates the convergence. 
Since the collaborative context largely narrows down the action space and
leads to a better policy solution in the sense of both effectiveness and efficiency, 
we can conclude that coordination based on collaborative context is effective. 
Also, comparing the performances of cA2C
and cA2C-v2 (red/green curves in Figure~\ref{fig:contextembed} (a)), apparently 
the policy context embedding (considering both geographic and collaborative context) 
is essential to performance, which greatly reduces the redundant policy search.

\subsection{Ablation study on grouping the locations}
This section studies the effectiveness of our regularization design for
LP-cA2C. One key difference between our work and traditional fleet management
works~\cite{godfrey2002adaptiveI,godfrey2002adaptiveII} is that we 
didn't assume the drivers in one location can only pick up the orders 
in the same location. On the contrary, one agent can also serve the orders
emerged in the nearby locations, which is a more realistic and complicated setting. 
% In this case, we need to consider a group of nearby locations instead of 
In this case, we regularize the number of agents repositioned into a set of
nearby grids close to the number of estimated orders at next time step.
This grouping regularization in \eq{\ref{eq:linearprog}} is more efficient than the
regularization in \eq{\ref{eq:linearprognoadj}} requiring the number of agents repositioned into each grid
is close to the number of estimated orders at that gird since lots of 
reposition inside the same group can be avoided. 
As the results shown in Table~\ref{tb:lp_reg}, using the group regularization in \eq{\ref{eq:linearprog}} reallocates less agents while achieves same best performance 
as the one in \eq{\ref{eq:linearprognoadj}} (LP-cA2C').
\begin{align}
\label{eq:linearprognoadj}
 \max_{\vy(\vs_t)} \hspace{1em}  &  (\vv(\vs_t)^T \vA_t - \vc_t^T)\vy(\vs_t) - \lambda(\vo_t - \vA_t \vy(\vs_t))^2 
\end{align}

\begin{table}[h!]
\scriptsize
\centering
\vspace{-0.1in}
\caption{The effectiveness of group regularization design. The results are averaged over three runs. }
\vspace{-0.1in}
\hspace{-1mm}
\scalebox{0.95}{
\begin{tabular}{|c|c|c|c|c|} \hline
       	 	 & Normalized GMV & ORR & Repositions &ROI \\ \hline\hline
LP-cA2C  	 & $113.56 \pm 0.61$ & $95.24\% \pm 0.40\%$ & $341774$ & $3.9663$ \\ \hline
LP-cA2C'      & $113.60 \pm 0.56$ & $95.27\% \pm 0.36\%$ & $304752$ & $4.4633$ \\ \hline
\end{tabular}
}
\label{tb:lp_reg}
\end{table}

\subsection{Qualitative study}

\begin{figure} 
% \vspace{-0.1in}
\begin{tabular}{c c}
\hspace{-5mm}\includegraphics[width=0.25\textwidth]{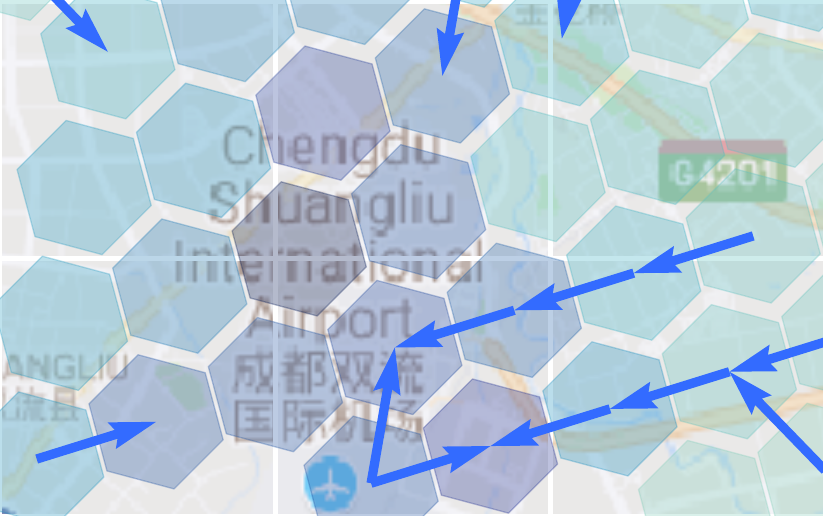} &
\hspace{-2.5mm}\includegraphics[width=0.25\textwidth]{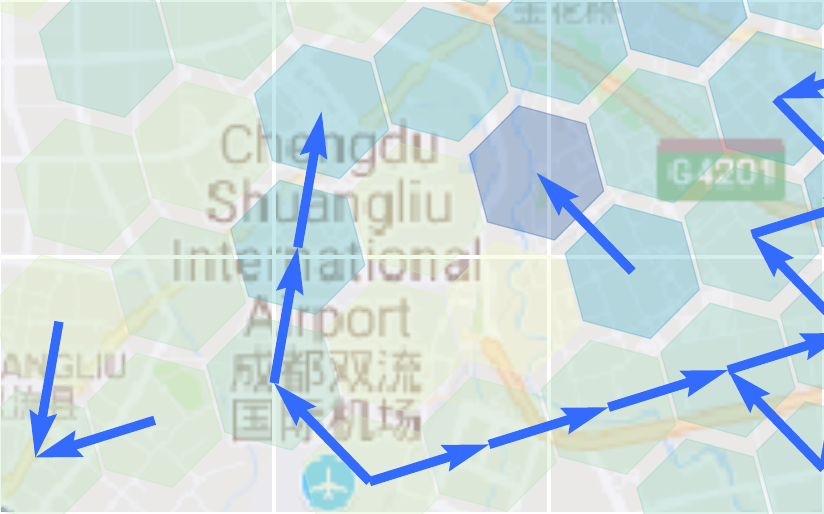}\\ 
\hspace{-4mm}(a) At 01:50 am. & (b) At 06:40 pm.
\end{tabular}
\caption{Illustration on the repositions nearby the airport at 1:50 am and 06:40 pm.
The darker color denotes the higher state value and the blue arrows denote the repositions.}
\label{fig:allocations} 
\end{figure}

\begin{figure}
% \vspace{-0.1in}
\centering
\includegraphics[width=0.45\textwidth]{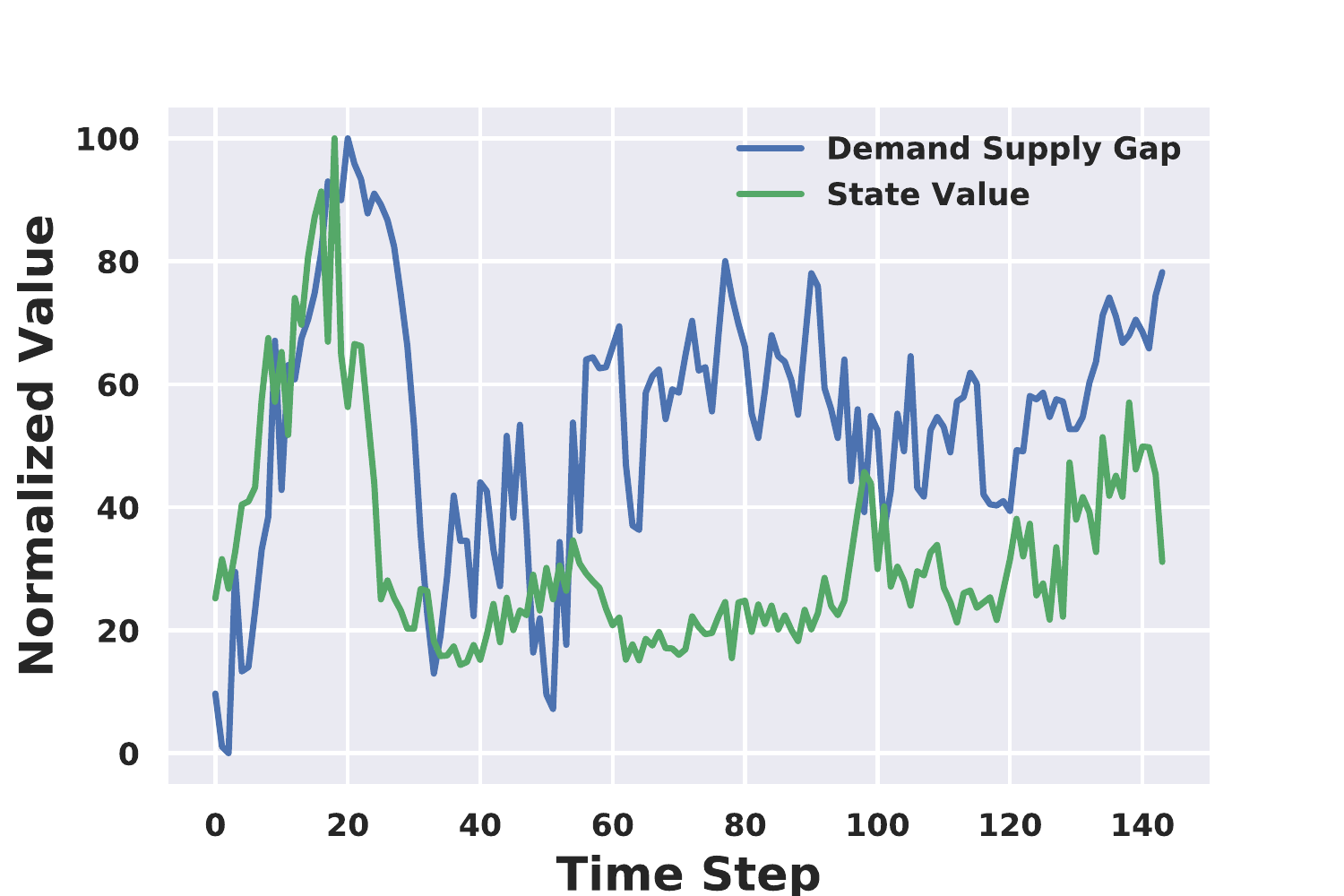}  
% \vspace{-0.1in}
\caption{The normalized state value and demand-supply gap over one day.}
\label{fig:valuefunction}
% \vspace{-0.2in}
\end{figure}

In this section, we analyze whether the learned value function can
capture the demand-supply relation ahead of time, and the rationality of
allocations. To see this, we present a case study on the region
nearby the airport.  The state value and allocation policy is acquired from
cA2C that was trained for ten episodes. We then run the well-trained cA2C on one
testing episode, and qualitatively exam the state value and allocations under the unseen dynamics. 
The sum of state values and demand-supply gap (defined as the number of orders minus the
number of vehicles) of seven grids that cover the CTU airport is visualized. 
% Similarly, we also plot
% the changes demand-supply gap (defined as the number of orders minus the
% number of vehicles) in that seven grids over one day. 
As seen in Figure~\ref{fig:valuefunction},
the state value can capture the future dramatic changes of demand-supply gap.
% It's worth noting that the current exact situation has
% not been seen by the value network and policy network in training stage, hence
% it shows the generalization ability of predicting state value under unseen
% dynamics. 
Furthermore, the spatial distribution of state values can be seen
in Figure~\ref{fig:allocations}. After the midnight, the airport has a large
number of orders, and less available vehicles, and therefore the state values of airport are
higher than other locations. During the daytime, more vehicles are available
at the airport so that each will receive less reward and the state values
are lower than other regions, as shown in Figure~\ref{fig:allocations}
(b). In Figure~\ref{fig:allocations} and Figure~\ref{fig:valuefunction}, we
can conclude that the value function can estimate the relative shift
of demand-supply gap from both spatial and temporal perspectives.
It is crucial to the performance of cA2C since the coordination is built upon
the state values. Moreover, as illustrated by blue arrows in Figure~\ref{fig:allocations},
we see that the allocation policy gives 
consecutive allocations
from lower value grids to higher value grids, which
can thus fill the future demand-supply gap and increase the GMV.

% Illustrate value function on the map. 
% Illustrate the reallocating actions. 
% combine those two figure in one map, like DQN~\cite{mnih2015human} Figure 1. 
% continuous
% reallocation action. reposition available vehicles to the high value region. 

% !TEX ROOT = main.tex

\section{Conclusions}
\label{sec:conclusion}
In this paper, we first formulate the large-scale fleet management problem
into a feasible setting for deep reinforcement learning.  Given this setting,
we propose contextual multi-agent reinforcement learning framework, in which
two contextual algorithms cDQN and cA2C are developed and both of them achieve
the large scale agents' coordination in fleet management problem. cA2C enjoys
both flexibility and efficiency by capitalizing a centralized value network
and decentralized policy execution embedded with contextual information.  It
is able to adapt to different action space in an end-to-end training
paradigm. A simulator is developed and calibrated with the real data provided
by Didi Chuxing, which served as our training and evaluation platform.
Extensive empirical studies under different settings in simulator have
demonstrated the effectiveness of the proposed framework.

% \IEEEraisesectionheading{\section{Introduction}\label{sec:introduction}}

\appendices
% \section{Proof of the First Zonklar Equation}
% Appendix one text goes here.

% % you can choose not to have a title for an appendix
% % if you want by leaving the argument blank
% \section{}
% Appendix two text goes here.

% use section* for acknowledgment
\ifCLASSOPTIONcompsoc
  % The Computer Society usually uses the plural form
  \section*{Acknowledgments}
\else
  % regular IEEE prefers the singular form
  \section*{Acknowledgment}
\fi

This material is based in part upon work supported by the National
Science Foundation under Grant IIS-1565596, IIS-1615597,
IIS-1749940 and Office of Naval Research N00014-14-1-0631, N00014-
17-1-2265.

% Can use something like this to put references on a page
% by themselves when using endfloat and the captionsoff option.
\ifCLASSOPTIONcaptionsoff
  \newpage
\fi

\newpage
\bibliographystyle{IEEEtran}
\bibliography{sample-bibliography}

% !TEX ROOT = main.tex

% \begin{IEEEbiographynophoto}{Kaixiang Lin} 
% \begin{figure}
% \includegraphics[width=0.15\textwidth]{fig/photo_Kaixiang.jpg}
% \end{figure}
% \end{IEEEbiographynophoto}

\vspace{-2em}
\begin{IEEEbiography}[{\includegraphics[width=1in, clip,keepaspectratio]{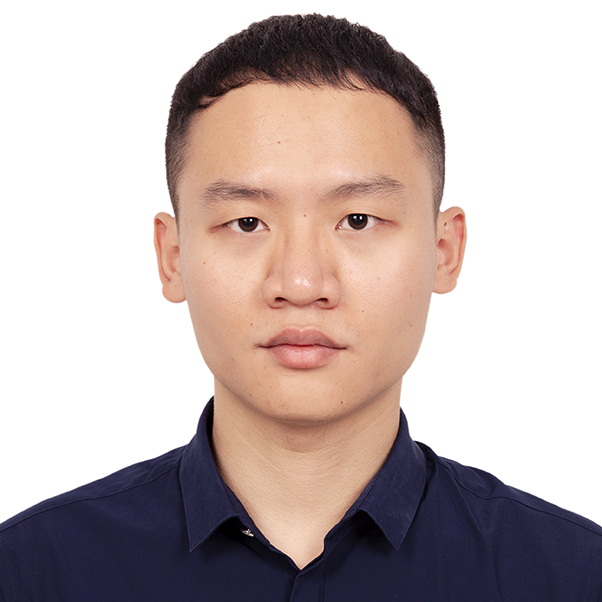}}]{Kaixiang Lin}
received the B.S. degree from the University
of Science and Technology of China, Hefei,
China, in 2014, and is currently working toward the
Ph.D. degree at Michigan State University, East Lansing,
MI, USA.
His research interests include machine learning,
data mining and reinforcement learning, with applications to the large-scale
traffic data and various domains.
\end{IEEEbiography}
\vspace{-1cm}
\begin{IEEEbiography}[{\includegraphics[width=1in,clip,keepaspectratio]{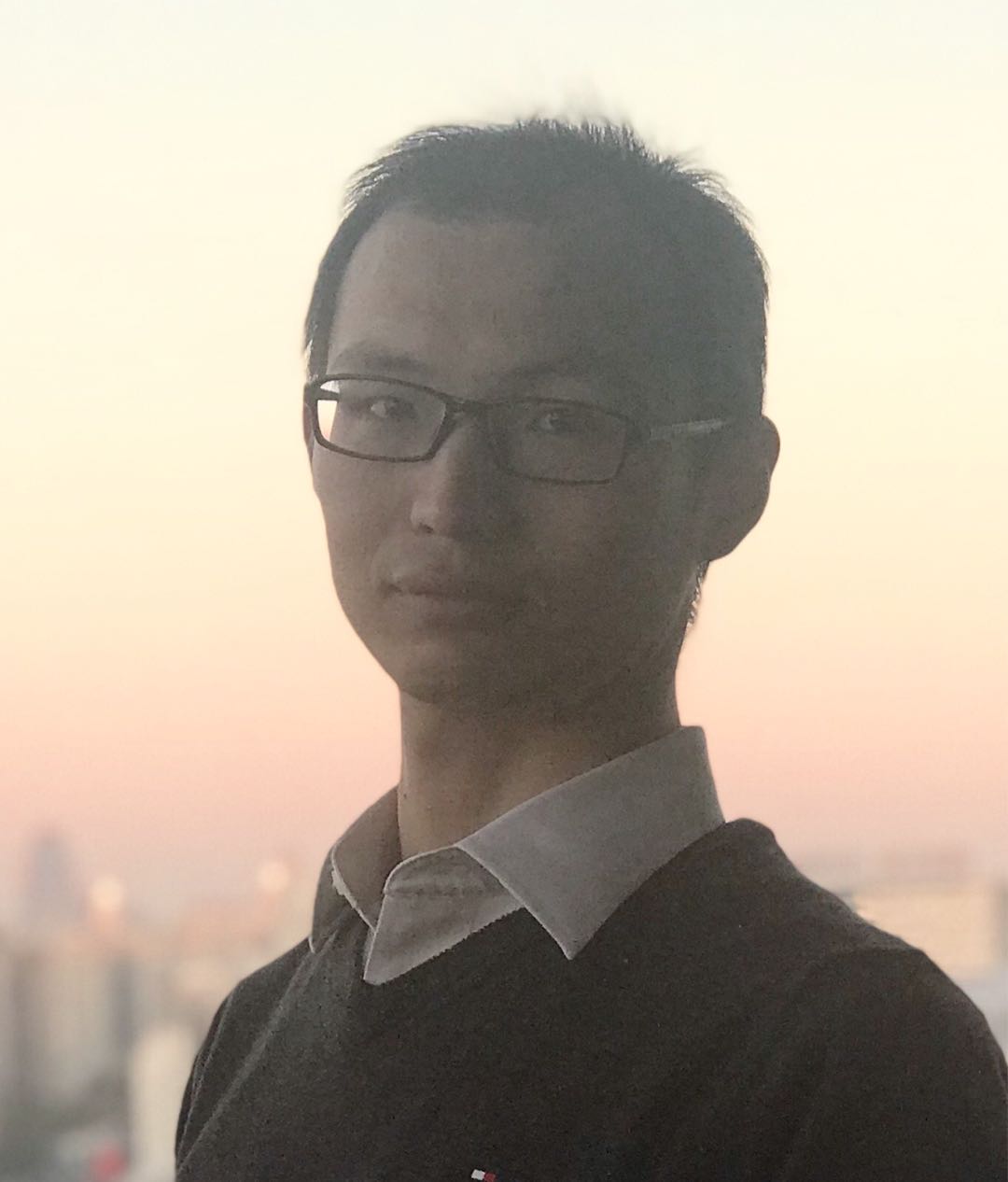}}]{Renyu Zhao} % Renyu Zhao % AI Labs of DiDiChuxing, Beijing, China 
received B.S. and M.E. degrees from China
Agriculture University, Beijing, China, and Peking University, Beijing, China,
in 2007 and 2013 respectively. He is a senior research engineer in Didi AI Labs, Beijing, China.
His research interests include spatiotemporal data mining,
stochastic process, as well as spectral graph theory, Bayesian analysis, etc.
\end{IEEEbiography}
\vspace{-1cm}

\begin{IEEEbiography}[{\includegraphics[width=1in,clip,keepaspectratio]{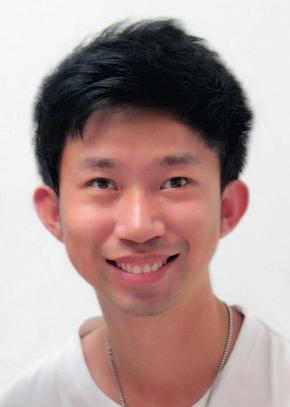}}]{Zhe Xu} received the B.S. degree in information
engineering from Shanghai Jiao Tong University,
Shanghai, China, in 2011. He received the Ph.D.
degree from the Institute of Image Communication and Networking, Shanghai
Jiao Tong University, Shanghai, China, and Quantum
Computation and Intelligent Systems, University of
Technology, Sydney, N.S.W., Australia.
His research interests include computer vision,
web mining, machine learning, and multimedia
analysis.
\end{IEEEbiography}
\vspace{-1cm}
\begin{IEEEbiography}[{\includegraphics[width=1in,clip,keepaspectratio]{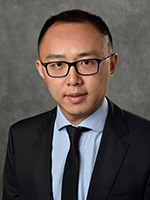}}]{Jiayu Zhou}
received the Ph.D. degree in computer
science from Arizona State University, Tempe, AZ,
USA, in 2014.
He is an Assistant Professor with the Department
of Computer Science and Engineering at Michigan
State University, East Lansing, MI, USA. His research
interests include large-scale machine learning
and data mining, and biomedical informatics.
Prof. Zhou served as Technical Program Committee
Member of premier conferences such as NIPS,
ICML, and SIGKDD. He was the recipient of the
Best Student Paper Award at the 2014 IEEE INTERNATIONAL CONFERENCE ON
DATA MINING (ICDM) and the Best Student Paper Award at the 2016 International
Symposium on Biomedical Imaging (ISBI).
\end{IEEEbiography}

% that's all folks
\end{document}